\documentclass[a4paper,11pt]{article}
\pdfoutput=1 

\usepackage{jcappub} 

\usepackage[T1]{fontenc} 
\usepackage{amsmath}
\usepackage{caption}
\usepackage{hyperref}
\usepackage[utf8]{inputenc} 
\usepackage{multirow}

\hypersetup{
    colorlinks=true,
    linkcolor=blue,   
    urlcolor=blue,
}

\title{Using CMB spectral distortions to distinguish between dark matter solutions to the small-scale crisis}

\author{James A.~D.~Diacoumis}
\author{and Yvonne Y.~Y.~Wong}

\affiliation{School of Physics, The University of New South Wales, Sydney NSW 2052, Australia}

\emailAdd{j.diacoumis@unsw.edu.au}
\emailAdd{yvonne.y.wong@unsw.edu.au}

\abstract{The dissipation of small-scale perturbations in the early universe produces a distortion in the blackbody spectrum of cosmic microwave background  photons.
In this work, we propose to use these distortions as a probe of the microphysics of dark matter on scales \(1 \, \textrm{Mpc}^{-1}\lesssim k \lesssim 10^{4} \, \textrm{Mpc}^{-1}\). 
We consider in particular models in which the dark matter is kinetically coupled to either neutrinos or photons until shortly before recombination, and compute the photon heating rate and the resultant \(\mu\)-distortion in both cases. 
 We show that the $\mu$-parameter is generally enhanced relative to $\Lambda$CDM for interactions with neutrinos, 
 and may be either enhanced or suppressed in the case of interactions with photons.  The deviations from the $\Lambda$CDM signal are potentially within the sensitivity reach of a PRISM-like experiment if \(\sigma_{\textrm{DM}-\gamma} \gtrsim 1.1 \times 10^{-30} \left(m_{\textrm{DM}}/\textrm{GeV}\right)  \textrm{cm}^{2}\) and \(\sigma_{\textrm{DM}-\nu} \gtrsim 4.8 \times 10^{-32} \left(m_{\textrm{DM}}/\textrm{GeV}\right)  \textrm{cm}^{2}\) for time-independent cross sections, and \(\sigma^{0}_{\textrm{DM}-\gamma} \gtrsim 1.8 \times 10^{-40} \left(m_{\textrm{DM}}/\textrm{GeV}\right)  \textrm{cm}^{2}\) and \(\sigma^{0}_{\textrm{DM}-\nu} \gtrsim 2.5 \times 10^{-47} \left(m_{\textrm{DM}}/\textrm{GeV}\right)  \textrm{cm}^{2}\) for cross sections scaling as temperature squared, coinciding with the parameter regions in which late kinetic decoupling  may serve as a solution to the small-scale crisis.
 Furthermore, these $\mu$-distortion signals differ from  those of warm dark matter (no deviation from $\Lambda$CDM) and  a suppressed primordial power spectrum (a strongly suppressed or negative $\mu$-parameter), demonstrating that CMB spectral distortion can potentially be used to distinguish between solutions to the small-scale crisis.}

\begin{document}
\maketitle
\flushbottom

\section{Introduction}
The canonical cold dark matter (CDM) paradigm has enjoyed remarkable success in its explanation of the universe's large-scale structure~\cite{Planck}. However, in spite of this success the CDM paradigm has a number of known issues pertaining to structure formation on small length scales. These issues include the ``missing satellites''~\cite{MissingSat1,MissingSat2} and ``too big to fail''~\cite{TBTF1,TBTF2} problems, in which the observed number of small satellite galaxies appears to be  fewer than predictions, 
and the ``core-cusp problem''~\cite{CuspCore1, CuspCore2, CuspCore3}, which refers to a discrepancy between the observationally inferred and the predicted dark matter halo density profiles in the inner parts of galaxies.

One popular solution to these small-scale issues is the warm dark matter (WDM) scenario, in which the dark matter is endowed with a small but non-negligible velocity dispersion, so that free-streaming can wash out structures below some velocity-dependent characteristic length scale~\cite{WDM1,WDM2,WDM3}.  In terms of the large-scale matter power spectrum, free-streaming damping in WDM cosmologies generically predicts a sharp cut-off of power at wavenumbers~$k$ larger than a free-streaming scale~$k_{\rm fs}$, where, in the case of a thermal relic WDM, $k_{\rm fs}$ is  related to the dark matter particle mass $m_{\rm DM}$ via 
$k_{\rm{fs}} \simeq 15.6 \left(m_{\rm DM}/1 \, \rm{keV}\right)^{4/3}\left(0.12/\Omega_{\rm{DM}}h^{2}\right)^{1/3} h\ {\rm Mpc}^{-1}$~\cite{LyA}.
 Observations of the Lyman-$\alpha$ (Ly$\alpha$) forest in quasar absorption spectra
 currently constrain  the thermal relic WDM particle mass to \(m_{\rm{WDM}} \gtrsim 5.2 \, \rm{keV} \, \,  (95\% \, CL)\) \cite{NewLyA}, corresponding to a free-streaming scale of approximately $100~{\rm Mpc}^{-1}$.

Another interesting possibility  is the late kinetic decoupling (LKD) scenario, wherein one couples the (nonrelativistic) dark matter to a relativistic ``heat bath'' (i.e., photons or standard model neutrinos) via elastic scattering until a fairly late time---up to the big bang nucleosynthesis (BBN) epoch or beyond---in the early universe~\cite{InteractingDM,InteractingDM2,LKD,LKD2,LKD3,Binder:2016pnr}.  Such an interaction 
forces the DM density perturbations on subhorizon scales to track those of the relativistic species when the coupling is strong, thereby preventing their growth.  Phenomenologically, the LKD matter power spectrum is  also characterised by a sharp cut-off of power  at large wavenumbers $k \gtrsim k_{\rm coll}$~\cite{Hofmann:2001bi}, albeit arising from collisional damping,  where the damping scale $k_{\rm coll} \propto \sqrt{\sigma_{{\rm DM}-X}/m_{\rm DM}}$ is set by the  elastic scattering cross section per dark matter mass.  Another notable feature  is the presence of small-scale dark acoustic oscillations, which, as predicted by linear perturbation theory, typically emerge at wavenumbers in the vicinity of~$k_{\rm coll}$~\cite{Loeb:2005pm}. 
Assuming  time-independent cross sections,
Milky Way satellites number counts currently constrain DM--photon scattering to $\sigma_{{\rm DM}-\gamma} \lesssim 3.7 \times 10^{-33} (m_{\rm DM}/{\rm GeV}) \ {\rm cm}^2$~\cite{Boehm:2014vja}, while a similar  $\sigma_{{\rm DM}-\nu} \lesssim 4 \times 10^{-33} (m_{\rm DM}/{\rm GeV}) \ {\rm cm}^2$~\cite{Wilknu}, derived from Ly$\alpha$ measurements, applies to DM--neutrino scattering.

Since by design both WDM and  LKD  replicate  CDM cosmology on the large length scales and possess the same gross phenomenology on the small, the question arises as to whether these solutions to the small-scale problems of CDM can ever be observationally distinguished from one another.  The challenge is especially formidable considering that potential tell-tale features, e.g., small-scale dark acoustic oscillations in LKD predicted by linear theory, generally occur at low redshifts on scales that have already undergone significant nonlinear evolution and  hence likely  suffer some degree of erasure.  Indeed,  collisionless $N$-body simulations of these alternative cosmologies appear to show small but inconclusive differences between WDM and LKD in their respective  dark matter halo properties~\cite{Schewtschenko:2014fca}.

In this paper we consider the potential of CMB spectral distortions as an alternative probe of dark matter physics on small scales.
Measurements of the CMB energy spectrum by the FIRAS instrument on COBE have shown that it is consistent with a perfect blackbody spectrum described by a temperature of \(T_{0} = 2.725 \pm 0.001\)~K~\cite{CMBTemp}.  Deviations from the blackbody spectrum are known as spectral distortions, and it has been long known that any (non-standard) energy release, e.g., from particle decays, that disturbs the thermodynamic equilibrium between photons and free electrons in the post-BBN universe ($z \lesssim 10^8$) could have sourced such distortions~\cite{Hu:1993gc}.   Null detection of $\mu$-type (chemical potential) and $y$-type (up-scattering) distortions by FIRAS has placed limits on the parameters
$|\mu| \lesssim 9 \times 10^{-5}$ and $|y| \lesssim 1.5 \times 10^{-5}$~(95\% C.L.)~\cite{COBE1,COBE2}, corresponding in both cases to a fractional change to the photon energy density of
$|\Delta \rho_\gamma/\rho_\gamma| \lesssim 6 \times 10^{-5}$ at  (\(95 \% \, \textrm{C.L.}\)).

 Interestingly, spectral distortion is also expected in standard $\Lambda$CDM cosmology as a consequence of spatial fluctuations and photon diffusion.
 Prior to recombination, primordial fluctuations combined with a tight coupling between photons and baryons engender acoustic oscillations in the primordial plasma  and hence local variations in the photon temperature  on subhorizon scales.  Concurrently, Thomson scattering enables  diffusion across  localities, causing photons of different temperatures to mix on scales comparable to the diffusion length at any given time.  Because such mixing effectively disturbs the local thermodynamic equilibrium of the photons, a small degree of spectral distortion is expected to arise as long as fluctuations are present in the photon fluid on scales  \({\cal O}(1) \,   \textrm{Mpc}^{-1} \lesssim k \lesssim {\cal O}(10^{4}) \,  \textrm{Mpc}^{-1}\) at $z \lesssim 10^8$~(see, e.g.,~\cite{Tashiro:2014pga,ChlubaRev} for a review);
for a standard power-law $\Lambda$CDM cosmology described by the Planck best-fit parameters, one may expect \(\mu \simeq 2 \times 10^{-8}\)~\cite{Chluba:2016bvg}.

While the FIRAS measurements are clearly not sufficiently sensitive to confirm or refute the power-law $\Lambda$CDM predictions, the sensitivities of proposed future experiments  such as PIXIE~\cite{PIXIE1, PIXIE2} are expected to improve to  \(\sigma(|\mu|) \sim 10^{-8}\) and \(\sigma(|y|) \sim 2\times10^{-9}\); the PRISM experiment could potentially push these numbers down by yet another order of magnitude~\cite{Andre:2013afa}.  This will not only enable a detection of the $\Lambda$CDM signal~\cite{Chluba:2012gq}, but the opportunity to probe those ``non-standard'' cosmologies that do not entail explicit energy injection at the homogeneous level, but nonetheless deviate from power-law $\Lambda$CDM on small length scales and alter the shape of the distortion through the aforementioned diffusion mechanism.
In this regard, WDM and LKD make excellent test subjects.

The paper is organised as follows.  We begin in section~\ref{sec:specdist} with a brief review of the physics of CMB spectral distortions, and estimate in section~\ref{sec:microphysics} the general impact of WDM and LKD microphysics on the distortion observables.  In sections~\ref{sec:dmnu} and~\ref{DMgamma} we model in detail the effects of DM--neutrino and DM--photon elastic scattering on the effective heating rate, and compute in particular the expected $\mu$-parameter.  We discuss the implications of our results for future experiments in section~\ref{sec:PRISM}.
Section~\ref{sec:conc} contains our conclusions.
All numerical estimates in this work assume cosmological parameter values given by the Planck 2015 vanilla $\Lambda$CDM best-fit~\cite{Planck}, and we use natural units, i.e., $ \hbar=c=1$, throughout the work.


\section{Spectral distortions}
\label{sec:specdist}

In the early universe, any release of energy inevitably creates a spectral distortion. If number-changing and energy-changing processes are both efficient, i.e., occur at a rate per photon much larger than the Hubble expansion rate, then the blackbody spectrum will be quickly restored after the momentary distortion. If  however these processes should become inefficient at any point and remain inefficient until the present time, the distortion may freeze in and  become observable. We briefly summarise some of the main features of spectral distortions below.
A detailed discussion  can be found in, e.g.,~\cite{Tashiro:2014pga,ChlubaRev}.

As a rule of thumb, at redshifts \(z \gtrsim 2\times 10^{6}\) thermalisation processes are extremely effective at erasing distortions, so that energy releases in this epoch generally do not survive to be detected.
When the redshift drops below \(z \sim 2\times 10^{6}\), the number-changing double Compton and Bremsstrahlung processes  begin to abate while  Compton scattering continues to redistribute the photon energy with efficiency.   This creates a condition in which a Bose--Einstein energy spectrum with a chemical potential can develop from an energy release.  Such a deviation from a Planck spectrum is called a $\mu$-distortion, and is quantified by $\Delta I_\nu = \mu M_\nu$, where $M_\nu$ is a characteristic shape  as a function of the photon frequency~$\nu$, and the $\mu$-parameter measures the degree of distortion.

 When, at \(z \lesssim 10^{4}\), even Compton scattering becomes too inefficient to thermally redistribute the photon energy, a $y$-distortion will form from an energy release.  Here, Compton scattering of photons on hot electrons effectively shifts low-energy photons in the Rayleigh--Jeans part of the spectrum to the high-energy Wien tail, creating a
 distortion $\Delta I_\nu = y \, Y_\nu$ whose characteristic shape $Y_\nu$ is distinctly different from the shape of a pure \(\mu\)-distortion, $M_\nu$, produced at an earlier time.  Thus, the shape of the spectral distortion carries in principle some degree of information on the time of the energy injection, with a pure $\mu$- and pure $y$-distortion representing respectively the early and the late extreme.  
It should be noted, however,  that while a \(\mu\)-distortion is uniquely associated with physics of the early universe, astrophysics at low redshifts, e.g., the Sunyaev--Zeldovich effect at  \(z \lesssim 20\), can also produce \(y\)-distortions that will in general swamp any such distortion produced around and prior to the epoch of recombination.

At intermediate times, \(10^{4} \lesssim z \lesssim 3\times 10^{5}\), when the universe's thermal conditions are in neither the $\mu$- nor the $y$-regime, the resultant distortion will be a superposition of the \(\mu\)-type and \(y\)-type extremes, plus an additional contribution generically labelled the \(r\)-distortion, i.e., 
\begin{equation}
\label{eq:distort}
\Delta I_\nu = y\, Y_\nu + \mu M_\nu +  R(\nu).
\end{equation}
 The exact shape of the \(r\)-distortion is governed by the energy injection history, and hence can in principle provide a useful handle on distinguishing between different scenarios of energy release in the early universe.
However, unlike the $\mu$- and $y$-distortions which have well-defined shapes,  the characterisation of the $r$-distortion is not unique; therefore, for simplicity, we shall not consider it in this work.

Lastly, we note that energy injection also shifts the overall temperature of the CMB photons.
This shift is however not a distortion to the blackbody spectrum, and therefore has not been included in equation~(\ref{eq:distort}).


\subsection{Effective heating rate} 
\label{blah}

A central quantity in the determination of spectral distortions is the effective heating rate of the CMB photons, ${\rm d} (Q/\rho_\gamma)/{\rm d} z$, defined as the 
 fractional energy release~$Q$ relative to the photon energy density~$\rho_\gamma=\rho_\gamma(z)$ as a function of redshift.  
For heating due to the presence of  primordial scalar fluctuations and diffusion in the photon fluid alone, this is given by~\cite{Chluba:2012gq,Khatri:2012rt}
\begin{multline} 
\label{master}
\frac{\textrm{d}\left(Q/\rho_{\gamma}\right)}{\textrm{d}z} = \frac{4a\dot{\kappa}}{\mathcal{H}}\int \, \frac{k^{2}\textrm{d}k}{2\pi^{2}}P_{\mathcal{R}}(k)\bigg[\frac{\left(3\Theta_{1}-v_{b}\right)^{2}}{3}+\frac{9}{2}\Theta_{2}^{2}\,  \\ - \frac{1}{2}\Theta_{2}\left(\Theta_{0}^{\textrm{P}}+\Theta_{2}^{\textrm{P}}\right) + \sum\limits_{\ell\geq3}(2\ell+1)\Theta_{\ell}^{2}\bigg].
\end{multline}
Here, \(\dot{\kappa} = a \sigma_{\textrm{T}}n_{e}\)
is the conformal Thomson scattering rate per photon, $\sigma_{\rm T}= 6.6524 \times 10^{-25} \, {\rm cm}^2$ the Thomson cross section, $n_e=n_e(z)$ the free electron number density, and
\(\mathcal{H} = a H(z)\) the conformal Hubble parameter.

Observe that the integrand is a product of the primordial curvature perturbations and transfer functions that describe how these perturbations evolve in the photon and the baryon fluids.  Specifically,  \(\Theta_{\ell}=\Theta_{\ell}(k,z)\) and \(\Theta_{\ell}^{\textrm{P}}=\Theta_{\ell}^{\textrm{P}}(k,z)\) denote respectively the transfer functions of the  \(\ell^{\textrm{th}}\) photon temperature and polarisation Legendre multipole moments, defined here via $\Theta (\vec{k},\hat{n}) = \sum_{\ell=0} (-{\rm i})^\ell (2 \ell + 1) \Theta_\ell(k) P_\ell (\hat{k} \cdot \hat{n})$, while \(v_{b}=v_b(k,z)\) is the transfer function of the baryon longitudinal velocity; these can  
 be computed numerically using a CMB Boltzmann solver such as the Cosmic Linear Anisotropy Solving System \texttt{CLASS}~\cite{CLASSI, CLASSII}.%
 \footnote{Available at \href{http://class-code.net/}{http://class-code.net/}} 
 Primordial fluctuations from inflation are encoded in the curvature power spectrum
 \(P_{\mathcal{R}}(k)\), commonly parameterised as a power law,
\begin{equation} 
\label{primordial}
P_{\mathcal{R}}(k) = \frac{2\pi^{2}}{k^{3}}A_{s}\left(\frac{k}{k_{0}}\right)^{n_{s} - 1-\frac{1}{2}n_{\textrm{run}}\textrm{ln}\left(\frac{k}{k_{0}}\right)},
\end{equation}
where \(n_{s}\) is the spectral index, \(n_{\textrm{run}}\) the running of the spectral index, \(A_{s}\) the amplitude, and \(k_{0}\) is the pivot scale. A number of previous works on CMB spectral distortions have put constraints on inflationary ``features'' in $P_{\cal R}(k)$  on the basis that such features alter the photon heating rate in the manner of equation~(\ref{master})~\cite{Primordial1, Primordial2, Cho:2017zkj}.  One of the  goals of the present work is to point out that spectral distortions can be equally sensitive to early universe physics that impact instead on the evolution of the transfer functions.

We focus on energy releases that take place at \(z \gtrsim 10^{4}\), which, for cosmologies consistent with current  observational data, is well before the recombination epoch.
At these redshifts the universe is dominated by radiation, and Compton/Thomson scattering keeps photons and baryons in a tightly-coupled fluid.
Tight coupling implies  \(v_{b} \simeq  3\Theta_{1}\), and higher-order photon temperature multipole moments  \(\Theta_{\ell\geq 2}\) become progressively smaller with~$\ell$.
Then, retaining terms up to $\ell=2$ and using the approximate solutions
 \(\Theta_{2}^{\textrm{P}} + \Theta_{0}^{\textrm{P}} \simeq (3/2) \Theta_{2}\) and \( \dot{\kappa} \Theta_{2} \simeq (8/15) k\Theta_{1} \)~\cite{HuSag1},
the effective heating rate~(\ref{master}) simplifies to~\cite{ChlubaGrin}
\begin{equation} \label{simp}
\frac{\textrm{d}\left(Q/\rho_{\gamma}\right)}{\textrm{d}z} \simeq \frac{4a }{\mathcal{H} \dot{\kappa}}\int \frac{\textrm{d}k}{2\pi^{2}}k^{4}P_{\mathcal{R}}(k)\frac{16}{15}\Theta_{1}^{2}.
\end{equation}
The dependence of the heating rate on subhorizon evolution  has thus been condensed into one single photon variable  \(\Theta_{1}\).

Equation~(\ref{simp}) may be further simplified using the approximate subhorizon solution~\cite{HuSag1}
\begin{equation} 
\label{Theta}
\Theta_{1} \simeq A \frac{c_s}{(1+R)^{1/4}} \, \textrm{sin}(kr_{s})\textrm{e}^{-k^{2}/k_{D}^{2}},
\end{equation}
valid for adiabatic initial conditions and wavenumbers~$k$ that enter the horizon well before recombination. Here, $R \equiv (3/4) \rho_b/\rho_\gamma   \ll 1$ is the baryon-to-photon energy density ratio,  $c_{s} =1/\sqrt{3(1+R)} $  the sound speed in the tightly-coupled photon--baryon fluid, \(r_{s}(z)= \int^\infty_z {\rm d} z' \, c_s a /{\cal H}\) the comoving sound horizon,  and \(k_{D}=k_D(z)\)~is the comoving diffusion damping scale given approximately by
\begin{equation} \label{diffusiondamping}
\partial_z k_D^{-2} = - \frac{c_s^2 a}{2 {\cal H} \dot{\kappa}} \left(\frac{16}{15} + \frac{R^{2}}{1+R}\right) .
\end{equation}
The prefactor~$A$ is the WKB mode amplitude on small scales, and is fixed by the fastest superhorizon growing mode to be
\begin{equation} \label{Amp}
A \simeq \left(1+\frac{4}{15}f_{\nu}\right)^{-1},
\end{equation}
where  \(f_{\nu} = \rho_{\nu}/(\rho_{\gamma} + \rho_{\nu}) \simeq 0.41\) is the ratio of the massless (free-streaming) neutrino energy density to the total energy density of relativistic species, and accounts for a small correction to the acoustic oscillation amplitude due to the presence of anisotropic stress in the neutrino fluid around horizon crossing.

 Ignoring baryon loading (i.e., formally setting $R=0$), the effective heating rate~(\ref{simp}) can now be expressed as
\begin{equation} \label{Cosmointegral}
\frac{\textrm{d}\left(Q/\rho_{\gamma}\right)}{\textrm{d}z} \simeq -8A^{2}\int \frac{k^{2} \, \textrm{d}k}{2\pi^{2}}  P_{\mathcal{R}}(k) \, \textrm{sin}^{2}(kr_{s}) \, k^{2}  \, \left[\partial_{z}k_{D}^{-2}\right] \, \textrm{e}^{-2k^{2}/k_{D}^{2}}.
\end{equation}
If \(P_{\mathcal{R}}(k)\) is smooth, then given $r_s \gg k_D^{-1}$ the oscillatory part of the integrand may be further averaged in $k$-space to obtain $\langle \sin^2(k r_s) \rangle  = 1/2$ and hence
\begin{equation} \label{fullsimplified}
\frac{\textrm{d}\left(Q/\rho_{\gamma}\right)}{\textrm{d}z} \simeq -4 A^2\int \, \frac{k^{2}\textrm{d}k}{2\pi^{2}}P_{\mathcal{R}}(k)\, k^2\,  \left[\partial_z k_D^{-2} \right] \, {\rm e}^{-2 k^2/k_D^2(k)}.
\end{equation}
This is the final form of the effective heating rate used in the public version of the cosmic thermalisation code~\texttt{CosmoTherm}~\cite{CosmoTherm1, ChlubaVis}.%
\footnote{Available at \href{http://www.cita.utoronto.ca/~jchluba/Science_Jens/CosmoTherm/Welcome.html}{www.Chluba.de/CosmoTherm}}

For an almost  scale-invariant curvature power spectrum (i.e., $n_s \simeq 1$, $n_{\rm run}=0$) it is now straightforward to establish that the integrand~(\ref{fullsimplified}) peaks at $k=k_{\rm D}/2 \simeq 2.0 \times 10^{-6} (1+z)^{3/2}$, where the numerical estimate here applies only during radiation domination.  This immediately translates to the sensitivity windows  $2 \ {\rm Mpc}^{-1} \lesssim k \lesssim 330 \ {\rm Mpc}^{-1}$ and $330 \ {\rm Mpc}^{-1} \lesssim k \lesssim  5700 \ {\rm Mpc}^{-1}$ over the redshifts $10^4 \lesssim z \lesssim 3 \times 10^5$ and $3 \times 10^5 \lesssim z \lesssim 2 \times 10^6$, respectively, to which the $\mu$/$y$-transition and the $\mu$-distortion eras correspond.


\subsection{Estimating the $\mu$- and the $y$-parameter}

Having specified the heating rate, the exact spectral distortion can be computed using the Green's function method~\cite{ChlubaVis}  that forms part of the thermalisation code~\texttt{CosmoTherm}.
For an estimate of the amounts of energy released that will eventually be observed as a \(\mu\)- and a y-distortion, however,  we may use
\begin{eqnarray} 
\label{distvisibility}
\frac{\Delta \rho_{\gamma}}{\rho_{\gamma}}\bigg|_{\mu} &=&  \frac{\mu}{1.401} \simeq \int^{\infty}_{0} \, \mathcal{J}_{\mu}(z)\mathcal{J}(z) \, \frac{\textrm{d}(Q/\rho_{\gamma})}{\textrm{d}z} \, \textrm{d}z , \\
 \label{distvisibility2}
\frac{\Delta \rho_{\gamma}}{\rho_{\gamma}}\bigg|_{y} & =& 4 y \simeq \int^{\infty}_{0} \, \mathcal{J}_{y}(z)\mathcal{J}(z) \, \frac{\textrm{d}(Q/\rho_{\gamma})}{\textrm{d}z} \, \textrm{d}z,\end{eqnarray}
where the $\mu$ and $y$ parameters parameterise respectively  the amplitude of distortion of the shape $M_\nu$ and $Y_\nu$ according to
 equation~(\ref{eq:distort}).

Here, ${\cal J} (z) \simeq {\rm e}^{-(z/z_{\rm dc})^{5/2}}$  is the distortion visibility function which quantifies the erasure of a distortion---primarily by double Compton scattering---between redshift~$z$ and the present day, 
where  \(z_{\textrm{dc}} \simeq 1.98 \times 10^6\) is the redshift at which double Compton scattering becomes inefficient; 
${\cal J}_{\mu}(z)$ and ${\cal J}_y(z)$ are the ``branching ratios'' of energy into $\mu$- and $y$-distortions respectively, given  approximately by~\cite{ChlubaVis}
\begin{eqnarray}
\mathcal{J}_{\mu}(z) &\simeq& 1 - \textrm{exp}\left[-\left(\frac{1+z}{5.8\times10^4}\right)^{1.88}\right], \\
\mathcal{J}_{y}(z) &\simeq& \left[ 1 + \left(\frac{1+z}{6\times10^{4}}\right)^{2.58} \right]^{-1}.
\end{eqnarray}
Note that ${\cal J}_\mu(z)+ {\cal J}_{y}(z) \neq 1$ due to the leakage of energy especially in the $\mu$/$y$-transition era,  $10^4 \lesssim z \lesssim 3 \times 10^5$, where a spectral distortion typically does not morph into either a $\mu$- or a $y$-distortion. This is the residual $r$-distortion discussed at the beginning of section~\ref{sec:specdist}.


\section{Impact of dark matter microphysics}
\label{sec:microphysics}

We are primarily interested in the redshift windows $10^4 \lesssim z  \lesssim  3 \times 10^5$ and $3 \times 10^5 \lesssim z \lesssim 2 \times 10^6$,  corresponding respectively to the  $\mu$/$y$-transition and $\mu$-distortion  eras.%
\footnote{We do not consider the $y$-distortion era, for reasons that (a) any $y$-distortion produced by dark matter microphysics in the early universe will be swamped by astrophysical sources, and (b) at $z \sim 1000$, the photon diffusion scale is $k_D \sim 0.1\, {\rm Mpc}^{-1}$, which is already probed by CMB anisotropies measurements; any  dark matter microphysics that impact on these scales would have already been ruled out.}
If dark matter microphysics should impact on the photon transfer functions in these windows, especially on scales $2 \ {\rm Mpc}^{-1} \lesssim k \lesssim 330 \ {\rm Mpc}^{-1}$ and 
$330 \ {\rm Mpc}^{-1} \lesssim k \lesssim 5700 \ {\rm Mpc}^{-1}$, then one would generically expect spectral distortions that differ from the power law $\Lambda$CDM predictions.  

Two observations are in order.   Firstly, while spectral distortions arise at $z \lesssim 10^6$, the wavenumbers to which distortions are primarily sensitive have been subhorizon
 typically since a much earlier time.  Using the criterion $k = {\cal H} \simeq 2 \times 10^{-6} (1+z) \ {\rm Mpc}^{-1}$, where the numerical estimate applies only during radiation domination, 
 horizon crossing takes place at $10^6 \lesssim z \lesssim  10^8$ for  $2 \ {\rm Mpc}^{-1} \lesssim k \lesssim 330 \ {\rm Mpc}^{-1}$,  
 while for $330 \ {\rm Mpc}^{-1} \lesssim k \lesssim 5700 \ {\rm Mpc}^{-1}$  we find $10^8 \lesssim z \lesssim  10^9$.
    Any modification to the photon transfer functions at these wavenumbers  from dark matter microphysics at such early times could in principle have survived down to the $\mu$-distortion and possibly the $\mu$/$y$-transition eras, even if by then this microphysics has ceased to be effective.  Thus, for the kind of physics considered in this work, the effective redshift window probed by spectral distortions in fact extends  to $ z \sim 10^9$.

Secondly, at $z \gtrsim 10^4$ the universe is radiation-dominated, so that  evolution of the metric perturbations is predominantly sourced by fluctuations  in the relativistic fluids, i.e., the photons and the neutrinos.  This means that any dark matter microphysics that alters only fluctuations in the nonrelativistic matter content can have but a negligible impact on the photon transfer functions unless the dark matter has a non-gravitational coupling to the relativistic species.  Thus, on this basis we can conclude without  further calculations that the $\mu$-distortion and $\mu$/$y$-transition signatures of WDM and CDM must be to first approximation identical.

In contrast, LKD scenarios explicitly couple the dark matter to a relativistic species~$X= \gamma,\nu$ via elastic scattering, and can thus be expected to alter the photon transfer functions on scales probed by spectral distortions if the relevant scattering processes are operational for some time at $z \lesssim 10^9$.
  Quantitatively this means the conformal scattering rate per $X$-particle, defined as
\begin{equation}
\dot{\mu}_X \equiv a \sigma_{{\rm DM}-X} n_{\rm DM}, 
\end{equation}
where $\sigma_{{\rm DM}-X}$ is the scattering cross section and $n_{\rm DM}=n_{\rm DM}(z)$  the dark matter number density,  should satisfy $\dot{\mu}_X \gtrsim {\cal H}$ for some duration between $z \sim 10^4$ and $z\sim  10^9$.  Assuming a constant cross section this is equivalent to requiring $\sigma_{{\rm DM}-X}$ to satisfy approximately 
\begin{equation}
\label{eq:constantx}
\left(\frac{\sigma_{{\rm DM}-X}}{{\rm cm}^2} \right)
\left(\frac{\rm GeV}{m_{\rm DM}} \right)  \gtrsim 
\begin{cases}
2 \times 10^{-33},  & \qquad \mu \ {\rm distortion} \\
 2 \times 10^{-32}, & \qquad \mu/y \ {\rm transition}
 \end{cases}
\end{equation}
where $m_{\rm DM}$ is the dark matter mass, in order for the scattering to impact on the $\mu$-distortion and $\mu$/$y$-transition eras respectively; the requirement tightens to $\gtrsim 2 \times 10^{-28}$ if the $\dot{\mu}_X \gtrsim {\cal H}$ condition is to be satisfied all the way down to $z \sim 10^4$.
If the cross sections scale with temperature squared,  i.e., $\sigma_{{\rm DM}-X} = \sigma_{{\rm DM}-X}^0 (T/T_0)^2$,
the corresponding requirements on their present-day values  are
 \begin{equation}
\label{eq:t2x}
 \left(\frac{\sigma_{{\rm DM}-X}^0}{{\rm cm}^2} \right)
\left(\frac{\rm GeV}{m_{\rm DM}} \right)   \gtrsim 
\begin{cases}
2 \times 10^{-51},  & \qquad \mu \ {\rm distortion} \\
 2 \times 10^{-48}, & \qquad \mu/y \ {\rm transition}
 \end{cases}
\end{equation}
and $\gtrsim 2 \times 10^{-36}$ for the entire duration of $10^4 \lesssim z \lesssim 10^9$.

\begin{table}[t]
\begin{center}
\begin{tabular}{lc|ccc|c}
\hline
\hline
& \multirow{2}{*}{$\sigma \propto T^n$} & \multirow{2}{*}{Planck+WP} & \multirow{2}{*}{MW satellites} & \multirow{2}{*}{Lyman-$\alpha$} & FIRAS  \\
&&&&&{\footnotesize ($1  \lesssim m_{\rm DM}/{\rm keV} \lesssim 100$)} \\
\hline
\multirow{2}{*}{$\gamma$} &
$n=0$ & $\lesssim 1.6\times 10^{-30}$ \cite{Wilkphoton} & $\lesssim 4 \times 10^{-33}$ \citep{Boehm:2014vja} & n.a. & $\lesssim 10^{-33}$ \cite{ChlubaKam} \\
&$n=2$ & $ \lesssim 1.2 \times 10^{-39}$ \cite{Wilkphoton} & n.a. & n.a. & $\lesssim 10^{-46}$ \cite{ChlubaKam} \\
\hline
\multirow{2}{*}{$\nu$} &
$n=0$ & $\lesssim 6 \times 10^{-28}$ \cite{Wilknu} &  n.a. & $\lesssim  10^{-33}$ \cite{Wilknu} & unconstrained\\
& $n=2$&  $\lesssim 8 \times 10^{-40}$ \cite{Wilknu} & n.a. & $\lesssim  10^{-45}$ \cite{Wilknu} & unconstrained  \\
\hline
\hline
\end{tabular}
\end{center}
\caption{Current 95\% C.L.\ constraints on the present-day value of $\sigma_{{\rm DM}-X}/m_{\rm DM}$ for DM--photon and DM--neutrino elastic scattering in units of ${\rm cm}^2$/GeV from various observations: CMB temperature and polarisation anisotropies from Planck and WMAP respectively (Planck+WP), Milky Way satellite number counts (MW satellites), intensity power spectrum of the Lyman-$\alpha$ forest, and CMB spectral distortions (FIRAS).  We have collated constraints for both the case of a time-independent cross section ($n=0$) and one in which the cross section scales as temperature squared ($n=2$).  An entry of ``n.a.'' denotes a scenario that is in principle constrainable by the observation in question but the constraint is not available in the literature, while ``unconstrained'' implies the scenario cannot be constrained by the said means.  
Note that the FIRAS constraints are based on spectral distortions due to kinetic energy transfer from the photon bath to the DM sector in the homogeneous background, and apply to DM masses only in the range $1  \lesssim m_{\rm DM}/{\rm keV} \lesssim 100$.  See text for details.\label{tab:current constraints}}
\end{table}

Current bounds on DM--photon and DM--neutrino elastic scattering from various observations are shown in table~\ref{tab:current constraints}.
Comparing these constraints with the sensitivity estimates~(\ref{eq:constantx}) and~(\ref{eq:t2x}), it appears that spectral distortions may be able to probe LKD with a high degree of complementarity with anisotropy measurements.  
It is  nonetheless important to emphasise that we have arrived at the sensitivity estimates~(\ref{eq:constantx}) and~(\ref{eq:t2x}) based on causality arguments alone; it remains to be seen how strongly specific scenarios of dark matter microphysics imprint on the photon transfer functions.  In section~\ref{sec:dmnu} we consider the case of DM--neutrino elastic scattering, and DM--photon scattering in section~\ref{DMgamma}.

Lastly, note that in table~\ref{tab:current constraints} existing spectral distortion bounds from FIRAS on the DM--photon scattering cross section have been derived on the basis of direct transfer of kinetic energy from the photon bath to the DM sector in the homogeneous background induced by the interaction~\cite{ChlubaKam}.  Such a mechanism produces a {\it negative} $\mu$-distortion, and differs from the scenarios considered in this work,  which operate on the inhomogeneous level.  Given a fixed DM energy density, spectral distortions caused by kinetic energy siphoned from the photon bath are inherently more severe for small DM masses simply because the total kinetic energy in the DM--photon system must now be shared between a larger number of particles.  For this reason, the FIRAS bounds derived in~\cite{ChlubaKam} apply only to  DM masses falling below \(m_{\rm DM} \lesssim 100\)~eV.  (In addition, a lower limit of \(m_{\rm DM} \gtrsim 1\)~eV has been imposed on the region of validity  to ensure that the DM is already nonrelativistic at the beginning of the distortion epoch.)
No such spectral distortion constraint exists on the DM--neutrino scattering cross section, obviously because this interaction has no direct impact on the background photon distribution.


\section{Dark matter--neutrino scattering}
\label{sec:dmnu}

A coupled DM--(almost massless) neutrino system is essentially analogous to the familiar photon--baryon system.  In the tight coupling limit, the DM--neutrino fluid experiences acoustic oscillations damped by neutrino diffusion~\cite{GravCluster}.  Diffusion damping causes the DM density perturbations to lose power on small scales, which serves as the basis of LKD as a solution to the small-scale problems of CDM cosmology.  At the same time, binding neutrinos to DM prevents them from free-streaming, thereby enhancing their energy density perturbations at the expense of a reduced anisotropic stress.

Given the effective Lagrangian of an interaction it is straightforward (although time-consuming) to write down the model-specific time evolution equations for the DM--neutrino perturbations~\cite{Isabelle,ETHOS}.   These can be solved together with the equations of motion for other cosmological fluids using~\texttt{CLASS} to obtain the LKD-modified photon and baryon transfer functions required to compute the heating rate~(\ref{master}).
However, since we are only concerned with the gross effect of DM--neutrino elastic scattering on spectral distortions,  in the interest of clarity and brevity we shall forego the detailed approach but simply adapt the relevant equations of motion for the photon--baryon system to our problem. 

This amounts to modifying the massless neutrino Boltzmann hierarchy in the conformal Newtonian gauge to~\cite{Wilknu, InteractingDM}
 \begin{align}
\begin{split}
 \label{eq:nuhierarchy}
\dot{\delta}_\nu &= -\frac{4}{3} \theta_\nu + 4 \dot{\phi}, \\
\dot{\theta}_{\nu} &= k^{2}\psi + k^{2}\left(\frac{1}{4}\delta_{\nu} - \sigma_{\nu}\right) -\dot{\mu}_\nu \left(\theta_{\nu} - \theta_{\textrm{DM}}\right), \\
\dot{\sigma}_{\nu} &= \frac{4}{15}\theta_{\nu} - \frac{3}{10} k F_{\nu 3} - \frac{9}{10}\dot{\mu}\sigma_{\nu}\\
\dot{F}_{\nu \ell} & = \frac{k}{2 \ell +1} \left[ \ell F_{\nu (\ell-1)} - (\ell +1) F_{\nu (\ell+1)} \right] -  \dot{\mu}_\nu F_{\nu \ell}, \qquad \ell \geq 3,
\end{split}
\end{align}
where, following the convention of~\cite{MaB}, $\delta_\nu=F_{\nu 0}$, \(\theta_{\nu}=(3/4) k F_{\nu 1}\) and $\sigma_\nu= F_{\nu2}/2$ are the neutrino energy density perturbation, velocity divergence and anisotropic stress respectively,  $F_{\nu \ell}$ the $\ell^{\rm th}$ neutrino Legendre multipole moment, \(\theta_{\textrm{DM}}\) the DM velocity divergence, $\psi$ and~$\phi$ perturbations in the line element ${\rm d} s^2 = a^2 [-(1+2 \psi) {\rm d}\eta^2 + (1-2 \phi) {\rm d}x^i {\rm d}x_i]$, and an overdot denotes partial differentiation with respect  to conformal time~$\eta$.  The DM--neutrino scattering is encoded in the terms proportional to $\dot{\mu}_\nu \equiv a \sigma_{{\rm DM}-\nu} n_{\rm DM}$, and in writing 
equation~(\ref{eq:nuhierarchy}) we have implicitly assumed that the scattering cross section~$\sigma_{{\rm DM}-\nu}$ has the same 
 angular dependence as Thomson scattering.
 
 The corresponding equations of motion for the dark matter perturbations are 
\begin{align} 
\begin{split}
\label{eq:dmeom}
\dot{\delta}_{\rm DM} & = - \theta_{\rm DM} + 3 \dot{\phi}, \\
\dot{\theta}_{\textrm{DM}} &= k^{2}\psi - \mathcal{H}\theta_{\textrm{DM}} - S_{\nu}^{-1}\dot{\mu}_\nu \left(\theta_{\textrm{DM}} - \theta_{\nu}\right),
\end{split}
\end{align}
where the presence of the factor \(S_{\nu} \equiv (3/4) \rho_{\textrm{DM}}/\rho_{\nu}\) stems from conservation of momentum in the coupled DM--neutrino system, 
and we have  omitted a pressure gradient term $k^2 c_{\rm DM}^2 \delta_{\rm DM}$ in the Euler equation, where $c_{\rm DM}$ denotes the DM sound speed. Typically, $c^2_{\rm DM} \sim T_\nu/m_{\rm DM} \ll 1/3$ when the DM is kinetically coupled to the neutrinos, i.e., $S^{-1}_\nu \dot{\mu}_\nu \gtrsim {\cal H}$, and decays away as $c^2_{\rm DM} \propto a^{-2}$  when $S^{-1}_\nu \dot{\mu}_\nu \lesssim {\cal H}$.
The said omission may impact on the detailed evolution of the DM perturbations at  wavenumbers $k \gtrsim {\cal H}/c_{\rm DM}$ in the weak coupling regime ${\cal H} \lesssim S^{-1}_\nu \dot{\mu}_\nu \lesssim k$, when the neutrino velocity perturbation $\theta_\nu$ becomes increasingly ineffective at prevailing over the ``intrinsic''  properties of the DM system to drive its dynamics.
Its impact on the neutrino perturbations (and ultimately the photon perturbations) is however negligible.

Note that in adopting equations~(\ref{eq:nuhierarchy}) and~(\ref{eq:dmeom}) to describe the DM--neutrino system we have implicitly assumed (a)~``massless'', i.e., ultrarelativistic neutrinos,  and (b)~the ``Thomson limit'' for the DM--neutrino interaction, i.e., scattering alters only the direction of a neutrino but not its energy, which requires that the average neutrino energy satisfies $\langle E_\nu \rangle \ll m_{\rm DM}$.
Given $4 \ {\rm eV} \lesssim \langle E_\nu \rangle \simeq 3.15 \times (4/11)^{1/3}T  \lesssim 400 \ {\rm keV}$ in the redshift window $10^4 \lesssim z \lesssim 10^9$ and an upper limit of $\sum m_\nu \lesssim 0.23$~eV on the neutrino mass sum~\cite{Planck}, we see that these assumptions are easily justified for dark matter masses in excess of $\sim 1$~GeV.

Lastly, because the new interaction is most effective at early times, it affects the evolution of perturbations on both superhorizon and subhorizon scales.  To ensure that  numerical solutions are free of transients one could either set the initial conditions analytically to the tracking solutions, or push the initialisation time back to an earlier epoch, giving the transients time to decay away.  While the first approach is standard in $\Lambda$CDM-type cosmologies, analytical tracking solutions are not readily available in cosmologies with new particle interactions.  Therefore, in this work we adopt the second approach, and take care to ascertain that the tracking solution has been reached numerically well before a wavenumber crosses the horizon.


\subsection{Effects on the photon transfer functions}
\label{sec:nueffects}

\begin{figure}[t]
\begin{center}
    \includegraphics[width=0.8\textwidth]{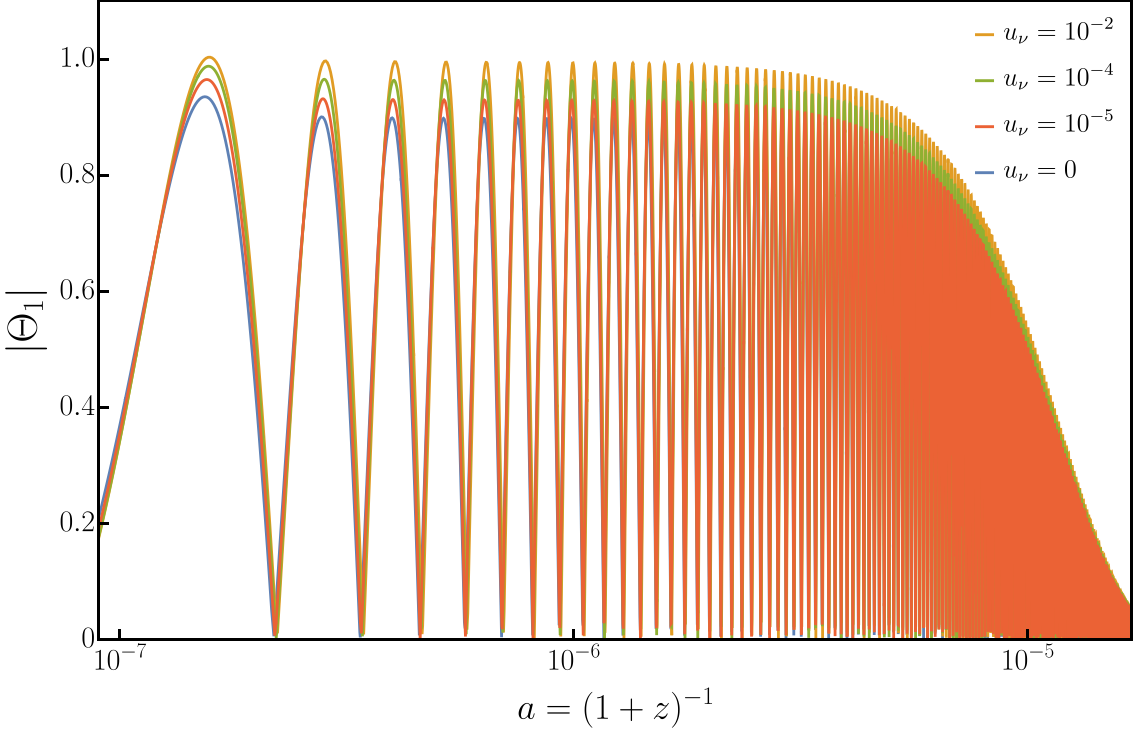}
        \includegraphics[width=0.8\textwidth]{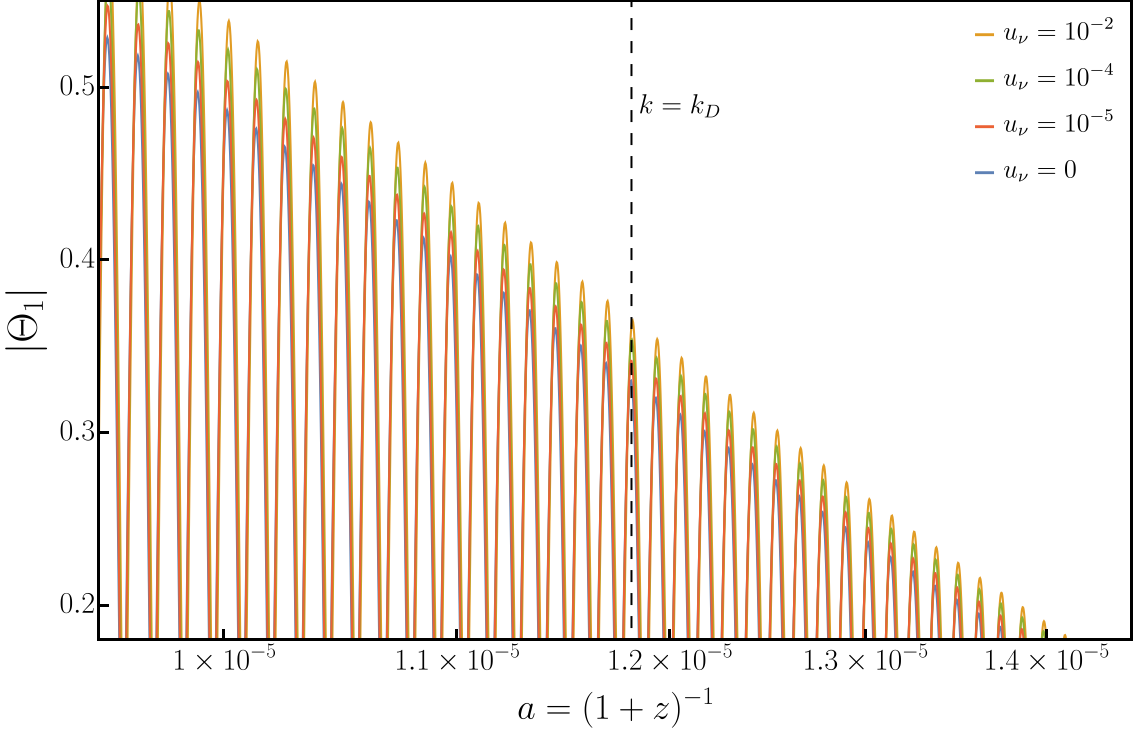}
    \caption{{\it Top}: Absolute value of the  photon temperature transfer function \(\Theta_{1}\)  at $k= 100\ {\rm Mpc}^{-1}$ as a function of the scale factor~\(a\) for a selection of time-independent DM--neutrino scattering cross sections: \(u_{\nu} = \{10^{-2},10^{-4}, 10^{-5},0\} \), where \(u_{\nu}\) is defined in equation~(\ref{nucross}).          {\it Bottom}: Same as the top panel, but zoomed in on the  diffusion damping epoch.
 The vertical dotted line denotes the time at which \( k = k_{D}\), where \(k_{D}\) is the diffusion damping scale.}
       \label{NeutrinoFunctions}
\end{center}
\end{figure}

Figure~\ref{NeutrinoFunctions} shows the photon transfer function~$\Theta_1$ at $k = 100 \, \rm{Mpc}^{-1}$ as a function of the scale factor~$a$ computed using \texttt{CLASS} for a selection of DM--neutrino scattering rates, assuming a time-independent cross section.
Following~\cite{Wilknu} we have defined the dimensionless parameter
\begin{equation} 
\label{nucross}
u_{\nu} \equiv \frac{\sigma_{\textrm{DM}-\nu}}{\sigma_{\textrm{T}}} \left(\frac{100 \, {\rm  GeV}}{m_{\rm DM}} \right)
\end{equation}
to represent the DM--neutrino scattering rate, where $\sigma_{\rm T}$ is the Thomson cross section, and $\sigma_{{\rm DM}-\nu} \simeq 6.7 \times 10^{-27} u_\nu   (m_{\rm DM}/{\rm GeV}) \ {\rm cm}^2$. If instead the cross section is proportional to the temperature squared,  we may write \(u_\nu = u_{\nu}^0a^{-2}\), where \(u_\nu^0\) is the present-day value.

We discuss below the impact of the DM--neutrino scattering on~$\Theta_1$ in the $\mu$-distortion and $\mu/y$-transition eras. For a complete discussion of the effects of the interaction on the photon perturbations, including during matter domination, we refer the reader to~\cite{Wilknu}.

\paragraph{Enhanced oscillation amplitude}
For a given wavenumber~$k$, if the tight-coupling condition 
\begin{equation}
\label{eq:tight}
\dot{\mu}_\nu \gg k, {\cal H}
\end{equation}
is satisfied before horizon crossing, the $\ell \geq 2$ multipole moments in the neutrino Boltzmann hierarchy are quickly damped to zero by the scattering term, thereby erasing the neutrino anisotropic stress $\sigma_\nu$  such that $\psi=\phi$ at the wavenumber concerned.  Upon horizon crossing (i.e., $k \sim {\cal H}$) during radiation domination, the total absence of anisotropic stress enables the neutrino perturbations $\delta_\nu$ and $\theta_\nu$ to participate in acoustic oscillations in the same way as the photon perturbations, leading to, all other things being equal,  the maximum possible acoustic oscillation amplitude in $\Theta_0$ and $\Theta_1$.

\begin{figure}[t]
\begin{center}
    \includegraphics[width=0.8\textwidth]{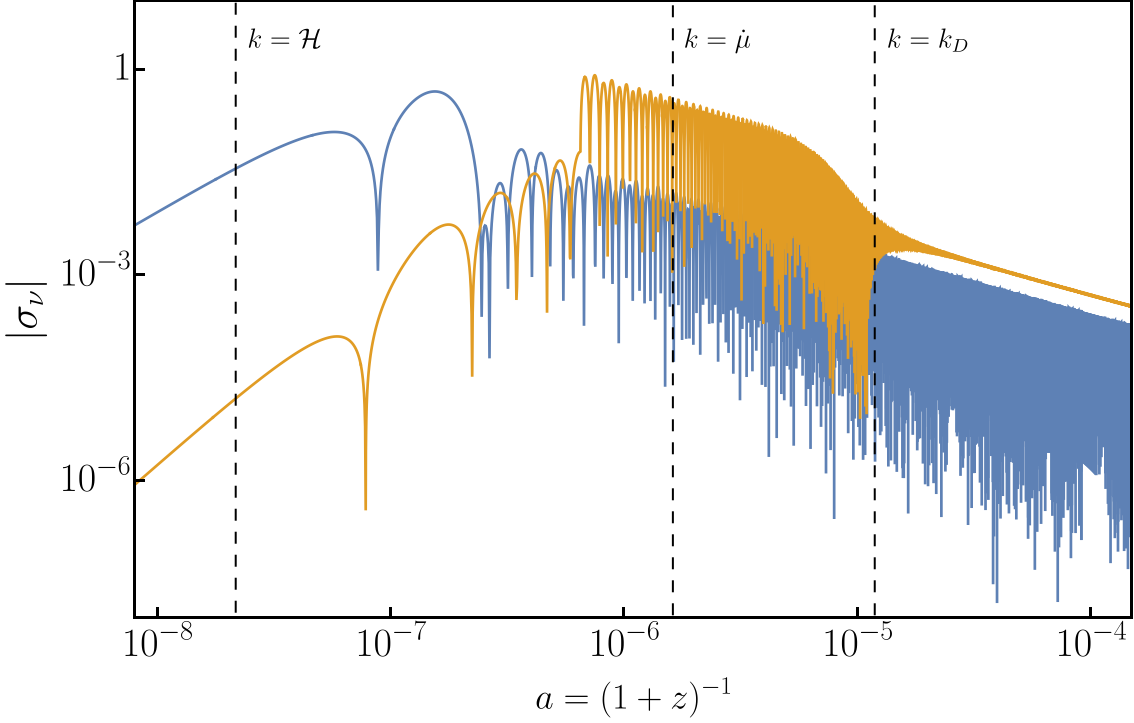}
    \caption{Absolute value of the neutrino anisotropic stress \(\sigma_{\nu}\)  at $k = 100\ {\rm Mpc}^{-1}$ as a function of the scale factor \(a\) for  \(u_{\nu} = 0\) (blue) and \(u_{\nu} = 1\times10^{-2}\)  (orange). 
     The vertical dotted lines denote, from left to right,  horizon crossing \(k = \mathcal{H}\), the end of the tight coupling epoch~\(k = \dot{\mu}_\nu\), and the onset of diffusion damping~\(k = k_{D}\).}
       \label{AnisotropicStress}
    \end{center}
\end{figure}

This limit is represented in figure~\ref{NeutrinoFunctions} by the $u_\nu=10^{-2}$ case (for $k = 100 \, \rm{Mpc}^{-1}$). The subhorizon evolution of $\Theta_1$ is numerically well approximated by equation~(\ref{Theta}), where the WKB amplitude~(\ref{Amp}) in the total absence of neutrino anisotropic stress at horizon crossing corresponds to setting $f_\nu=0$ so that $A \simeq 1$.  Figure \ref{AnisotropicStress} shows the corresponding neutrino anisotropic stress, which is clearly significantly lower than its standard $\Lambda$CDM (i.e., \(u_{\nu} = 0\)) counterpart around horizon crossing.

If conversely superhorizon evolution of the neutrino perturbations is characterised by the opposite of condition~(\ref{eq:tight}), i.e.,
\begin{equation}
\label{eq:nottight}
\dot{\mu}_\nu \ll k, {\cal H},
\end{equation}
 then we simply recover the standard free-streaming neutrino case, where the presence of anisotropic stress at horizon crossing attenuates the amplitude of the subsequent acoustic oscillations; for the canonical $f_\nu \simeq 0.41$ the WKB amplitude~(\ref{Amp}) evaluates to $A\simeq 0.9$.  In the intermediate regime where neither condition~(\ref{eq:tight}) nor~(\ref{eq:nottight}) can be satisfied---as represented in figure~\ref{NeutrinoFunctions} by $u_\nu = 10^{-3},10^{-4}$---the oscillation amplitude falls between $A\simeq 0.9$ and $\simeq 1$.  

It is also interesting to note that given the assumed time dependence of the scattering cross sections (constant or scaling with $T^2$), a wavenumber~$k$ that initially satisfies the tight coupling condition~(\ref{eq:tight}) will eventually reach the opposite limit~(\ref{eq:nottight}) after horizon crossing, providing an opportunity for the neutrino anisotropic stress to be regenerated subhorizon from~$\delta_\nu$ and~$\theta_\nu$.   The regeneration effect can be seen in figure \ref{AnisotropicStress}, where in the $u_\nu=10^{-2}$ case 
the anisotropic stress receives a significant boost around  \(k \sim \dot{\mu}_\nu \). We have verified that this boost has no discernible effect on \(\Theta_{1}\) (see bottom panel of figure~\ref{NeutrinoFunctions}), which can be understood as follows. 

The subhorizon dynamics of photon perturbations in the tightly-coupled limit is essentially that of a driven harmonic oscillator described by an approximate equation of motion (for the temperature monopole $\Theta_0$)~\cite{HuSag1}
\begin{equation} \label{oscillations}
\ddot{\Theta}_{0} + \frac{\dot{R}}{(1+R)}\dot{\Theta}_{0} + k^{2}c^{2}_{s}\Theta_{0} = \ddot{\phi} + \frac{\dot{R}}{(1+R)}\dot{\phi} - \frac{k^{2}}{3}\psi,
\end{equation}
where the gravitational potentials act as the ``driving force''. The neutrino anisotropic stress enters the picture via the difference between the potentials
\begin{equation}
k^{2}\left(\phi - \psi\right) =\frac{3}{2 M_{\rm pl}^2}  a^2 \sum_{i=\gamma,\nu} (\bar{\rho}_i + \bar{P}_i) \sigma_i \simeq 6\, {\cal H}^2 f_\nu \sigma_\nu,
\end{equation}
where  \(M_{\rm{pl}}\) is the reduced Planck mass, and we have assumed for the last approximate equality the photon anisotropic stress to be negligible in the tightly-coupled limit.
  Importantly, the potentials quickly decay away as soon as a wavenumber  crosses into the horizon.  This means that as a driving force for  photon acoustic oscillations, the presence or otherwise of neutrino anisotropic stress can only be of consequence around the time of horizon crossing, and even then the difference $\sigma_\nu$ engenders in $\Theta_0, \Theta_1$ is merely at the 10\% level.  Thus, we conclude that the subhorizon regeneration of $\sigma_\nu$ around $k \sim \dot{\mu}_\nu$ has no impact on the dynamics of the photon perturbations.

\paragraph{Peak shift}

If the tight-coupling condition~(\ref{eq:tight}) continues to be satisfied  after horizon crossing, then the coupled DM--neutrino fluid has a sound speed given by \(c_{\textrm{DM}-\nu} = 1/\sqrt{3 (1+ S_\nu)}\), which  is always smaller than the sound speed of the photon--baryon fluid because $S_\nu > R$. This mismatch is communicated via the metric perturbations~$\psi,\phi$ to the (driven) acoustic oscillations of the photon--baryon fluid for which $\psi,\phi$ serve as an external driving force, alters the effective sound horizon~\(r_{s}\), and ultimately manifests itself as a small ``peak shift'' in $k$-space in the oscillations of~\(\Theta_{1}\) at a given redshift.

While this peak shift can have important consequences for the CMB anisotropies (because there one actually measures the first few peak locations), for spectral distortions it is largely irrelevant.   At any given time in the spectral distortion era, the sound horizon~$r_s$ is typically much longer than the diffusion length scale~$k_D^{-1}$,  so that rapid oscillations of~$\Theta_1^2$  in $k$-space always average to $1/2$ within one diffusion length independently of the precise value of~$r_s$.   The peak shift therefore becomes essentially unobservable.


\subsubsection{Modelling the photon transfer functions}
\label{sec:numodel}

To compute the effective heating rate in the presence of DM--neutrino scattering in the redshift window of interest, we may take equation~(\ref{simp}) as a starting point, and feed in the appropriate photon transfer function~$\Theta_1$ outputted by a Boltzmann solver such as \texttt{CLASS}.  In practice, however, the highly oscillatory nature of the solution makes it difficult to track numerically at the kind of large wavenumbers, i.e., ${\cal O}(1)  \ {\rm Mpc}^{-1} \lesssim k \lesssim {\cal O}(10^4) \ {\rm Mpc}^{-1}$,  to which the heating rate is sensitive.  We therefore resort to approximate analytic solutions and interpolating functions.

As discussed above, for wavenumbers that enter the horizon during radiation domination, DM--neutrino scattering modifies primarily the amplitude of the photon--baryon acoustic oscillations by an amount that depends on the neutrino anisotropic stress present around horizon crossing ($k \sim {\cal H}$). The modifications are well approximated by the WKB amplitude~(\ref{Amp}) in the limits $\dot{\mu}_\nu \gg {\cal H}$ and $\dot{\mu}_\nu \ll  {\cal H}$, and can be equivalently summarised as
\begin{equation} 
\label{AmpCouple} 
A(k, u_\nu) \simeq
\begin{cases}
\left(1 + \frac{4}{15}f_{\nu}\right)^{-1} \simeq 0.9,& a_{\rm d} \ll a_{H} \\
\qquad 1, &  a_{\rm d} \gg a_{H} 
\end{cases}
\end{equation}
where 
\begin{equation}
a_{H}(k) \simeq \frac{H_0 \sqrt{\Omega_r}}{k} \simeq 2.2 \times 10^{-6} \left(\frac{{\rm Mpc}^{-1}}{k} \right)
\end{equation}
and 
\begin{equation}
a_{\rm d}(u_\nu) \simeq
\begin{cases}
3 H_0 M_{\rm pl}^2 \frac{\Omega_{\rm DM}}{\sqrt{\Omega_r}} \frac{\sigma_{{\rm DM}-\nu}}{m_{\rm DM}}  \simeq 0.012  \, u_\nu,  &\qquad \sigma_{{\rm DM}-\nu} = {\rm constant} \\
\left(3 H_0 M_{\rm pl}^2 \frac{\Omega_{\rm DM}}{\sqrt{\Omega_r}} \frac{\sigma_{{\rm DM}-\nu}^0}{m_{\rm DM}} \right)^{1/3} \simeq 0.23 \, {u_\nu^0}^{1/3},& \qquad \sigma_{{\rm DM}-\nu} \propto T^2 
\end{cases} 
\end{equation}
are defined, respectively, as the scale factors at which $k (a_H)= {\cal H}(a_H)$ (horizon crossing) and $\dot{\mu}_\nu(a_{\rm d}) = {\cal H}(a_{\rm d})$ (kinetic decoupling) are satisfied.
 Note that \(A(k,u_\nu)\) is \textit{not} a function of time; holding~$u_\nu$ fixed, 
 the two limits of~(\ref{AmpCouple}) are defined by the wavenumber~$k$ alone (up to the background cosmological parameters). 
 
In reality there should be a smooth transition between the two limiting regimes, where, for a fixed $u_\nu$, photon perturbations at intermediate wavenumbers receive
partial enhancements to their acoustic oscillation amplitudes.  We model this transition regime with an interpolating function in $x \equiv a_{\rm d}/a_{H}$
of the form $A(k,u_\nu)  =  (1+(4/15) f_{\nu} \, \exp [-(x/c_0)^{c_1}])^{-1}$.  Specifically, for the scattering cross sections in question,
\begin{align}
\begin{split}
 \label{NewAmp}
A(k,u_\nu) 
 & \simeq
 \begin{cases}
 \left(1+\frac{4}{15}f_{\nu} \, \textrm{exp}\left[-\left(\frac{k \, u_{\nu}}{7.82 \times 10^{-3} \, {\rm Mpc}^{-1}}\right)^{0.384}\right]\right)^{-1}, & \qquad \sigma_{{\rm DM}-\nu} = {\rm constant} \\
 \left(1+\frac{4}{15}f_{\nu} \, \textrm{exp}\left[-\left(\frac{k \, {u_{\nu}^0}^{1/3}}{1.02 \times 10^{-4} \, {\rm Mpc}^{-1}}\right)^{0.746}\right]\right)^{-1}, & \qquad \sigma_{{\rm DM}-\nu} \propto T^2
 \end{cases}
\end{split}
\end{align}
where the coefficients have been obtained from a least-square fit to a set of output from \texttt{CLASS} at scales $0.1 \leq k/{\rm Mpc}^{-1} \leq 100$, and for \(u_{\nu} = \{10^{-2},  10^{-3}, 10^{-4}, 10^{-5} \}\) in the time-independent case and \(u_{\nu}^0 = \{10^{-14},  10^{-16}, 10^{-17}, 10^{-18}, 10^{-19} \}\) for cross sections scaling as~$T^2$.
The rationale here is that all wavenumbers that cross the horizon in the post-BBN era under the same circumstances---in this instance, the same~$x$ value---evolve subhorizon in a self-similar way (modulo diffusion damping) until  the end of the radiation domination era.

Then, following the same arguments that lead to equation~(\ref{fullsimplified}), we find an effective heating rate  of
\begin{equation}
 \label{FinalHeating}
\frac{\textrm{d}\left(Q/\rho_{\gamma}\right)}{\textrm{d}z} \simeq -4 \int \, \frac{k^{2} \, \textrm{d}k}{2\pi^{2}} \, A^2(k,u_\nu)  \, P_{\mathcal{R}}(k) \, k^2 \, \left[\partial_{z} k_D^{-2}\right] \, \textrm{e}^{-2k^{2}/k_{D}^{2}}
\end{equation}
in the presence of DM--neutrino elastic scattering.  Observe that the amplitude squared is now inside the integral as it depends explicitly on~\(k\).


\subsection{Expected spectral distortions}

\begin{figure}[t]
\begin{center}
    \includegraphics[width=0.8\textwidth]{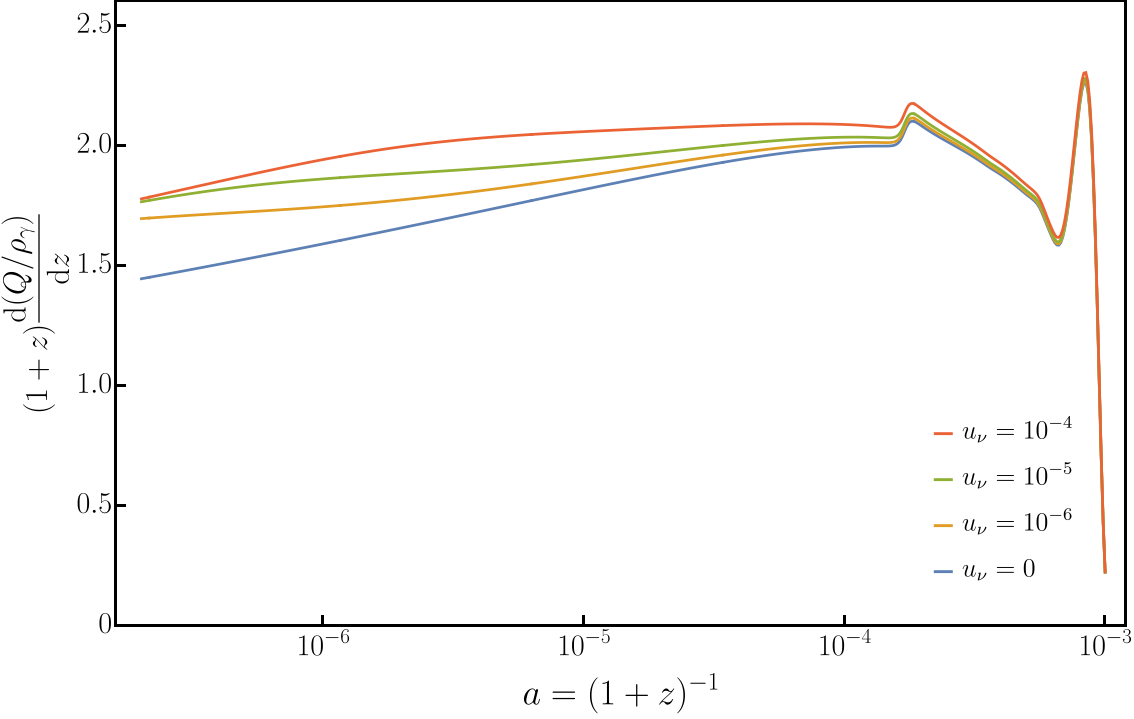}
    	\includegraphics[width=0.8\textwidth]{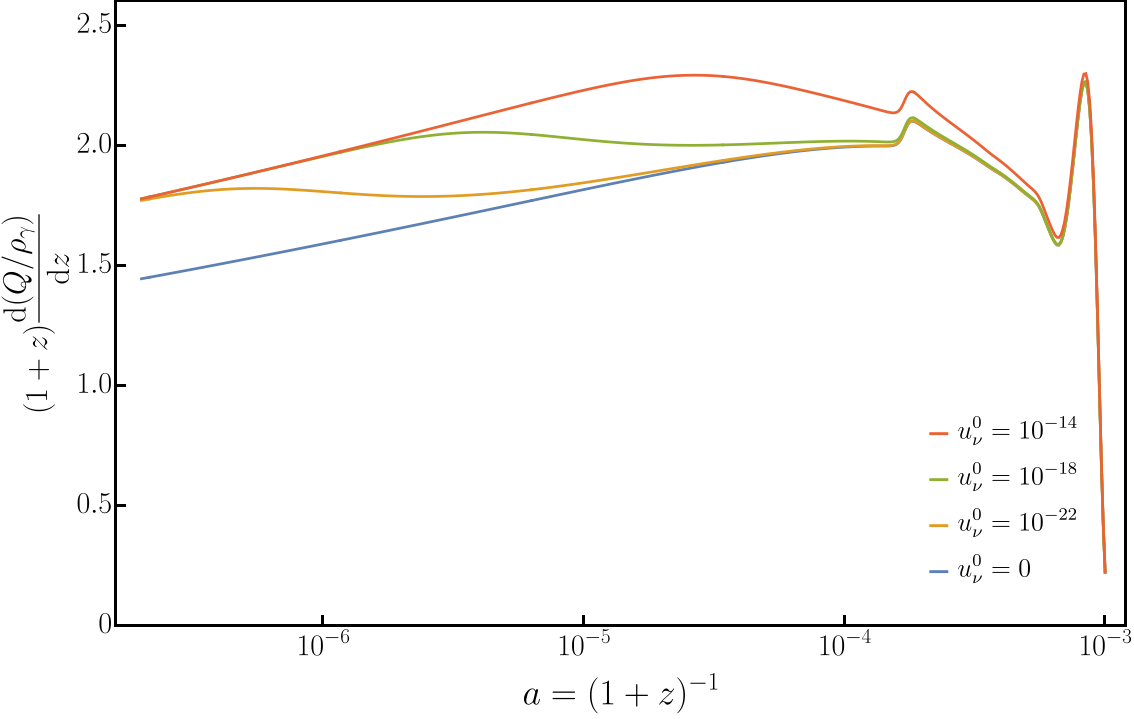}
    \caption{{\it Top}: Effective heating rate as a function of the scale factor \(a\) for a selection of DM--neutrino scattering cross sections: from top to bottom, $u_\nu=\{10^{-4}, 10^{-5}, 10^{-6},0 \}$.  In all cases the amplitude of the primordial power spectrum has been set to \(A_s = 1\) at the pivot scale \(k_{0} = 0.05 \, \textrm{Mpc}^{-1}\),
     the spectral index to \(n_{s} = 0.96\), and we assume no running \(n_{\textrm{run}} = 0\).  All other cosmological parameters assume the Planck 2015 vanilla best-fit values.  {\it Bottom}: Same as the top panel, but for  DM--neutrino scattering cross sections proportional to the temperature squared: from top to bottom, $u^{0}_{\nu}=\{10^{-14}, 10^{-18}, 10^{-22},0 \}$.}
       \label{NeutrinoHeating}
    \end{center}
\end{figure}

Figure~\ref{NeutrinoHeating} shows the heating rate for a selection of DM--neutrino scattering rates computed using~\texttt{CosmoTherm} incorporating the  modification~(\ref{FinalHeating}). The background cosmology is taken to be standard vanilla $\Lambda$CDM  specified by the best-fit values to the Planck 2015 data.

The time dependence of the heating rate in the presence of a nonzero $u_\nu$ can be understood as follows.
As discussed in section~\ref{blah}, the integrand~(\ref{FinalHeating}) is primarily sourced by modes at \(k \sim k_{D}/2 \simeq 2 \times 10^{-6} \, a^{-3/2}\). 
Accordingly, at early times, the integral is dominated by large $k$ modes which have received an enhancement because of the tightly-coupled DM--neutrino system at horizon crossing.  This results in an enhanced heating rate.
 In contrast, at later times the integral becomes dominated by small $k$ modes that have received little or no enhancement as they crossed into the horizon when decoupling between the DM and neutrinos had already occurred.   Thus, the heating rate approaches the \(\Lambda \rm{CDM}\) value.

\begin{figure}[t]
\begin{center}
    \includegraphics[width=0.9\textwidth]{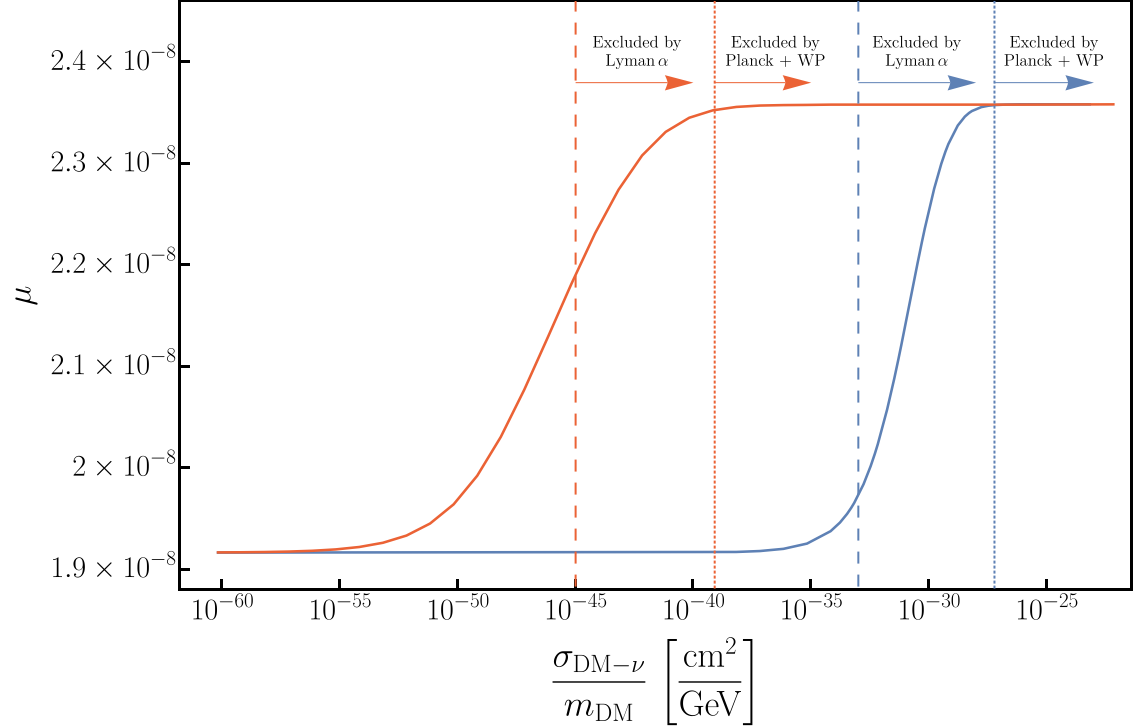}
    \caption{Expected \(\mu\)-parameter as a function of the  present-day DM--neutrino scattering cross section.  Here, the blue line denotes the case of a time-independent cross section, and the red line the case in which the cross section is proportional to the temperature squared. In both cases we assume a primordial power spectrum amplitude of \(A_{s} = 2.2\times10^{-9}\) at the pivot scale \(k_{0} = 0.05 \, \textrm{Mpc}^{-1}\), a spectral index  \(n_{s} = 0.96\), and no running  \(n_{\textrm{run}} = 0\). The vertical lines indicate the parameter regions presently excluded by various observations at 95\%~C.L.}
       \label{NeutrinoMu}
    \end{center}
\end{figure}

Figure~\ref{NeutrinoMu} shows the expected \(\mu\)-distortion as a function of the present-day value of $\sigma_{{\rm DM}-\nu}/m_{\rm DM}$, for time-independent cross sections as well as cross sections that scale as temperature squared.
In both cases the curves tend towards \(\mu = 1.91 \times 10^{-8}\)  in the limit \(\sigma_{{\rm DM}-\nu} \rightarrow 0\), recovering the $\Lambda$CDM prediction from dissipation of small scale perturbations. Note that this number is larger than that quoted in, e.g.,~\cite{Chluba:2012gq}, by \(2.7\times 10^{-9}\), which can be attributed to a partial cancellation by a negative $\mu$-distortion due to the Bose--Einstein condensation of the CMB photons~\cite{BEC}.

As we increase $\sigma_{{\rm DM}-\nu}/m_{\rm DM}$, evidently an enhanced \(\mu\)-distortion can result from the DM--neutrino coupling irrespective of the exact time dependence of the scattering cross section.
 However, the enhancement always saturates at about 20\% of the base $\Lambda$CDM value, which is a direct consequence of  the fact that removing neutrino anisotropic stress at horizon crossing can only raise at maximum the photon acoustic oscillation amplitude by about 10\%, as shown in section~\ref{sec:nueffects}. 
  The implications of this enhancement for future experiments will be discussed in section~\ref{sec:PRISM}.

\section{Dark matter--photon scattering}
\label{DMgamma}

The case of DM--photon elastic scattering may likewise be modelled by a simple modification of the equations of motion in the photon and the dark matter sectors in analogy with Thomson scattering.   In the photon sector, this entails additional collision terms proportional to the conformal DM--photon scattering rate~$\dot{\mu}_\gamma=a \sigma_{{\rm DM}-\gamma} n_{\rm DM}$ in the collision integrals of the Boltzmann equations:
\begin{eqnarray}
\label{eq:collisionintegralT}
\left( \frac{\partial F_\gamma}{\partial \eta} \right)_C & = & \dot{\kappa} \left[ -F_\gamma + F_{\gamma 0} + 4 \hat{n} \cdot \vec{v}_b - \frac{1}{2} \left( F_{\gamma 2} + G_{\gamma 0} + G_{\gamma 2}\right) P_2\right]  \\
&& +\dot{\mu}_\gamma \left[ -F_\gamma + F_{\gamma 0} + 4 \hat{n} \cdot \vec{v}_{\rm DM} - \frac{1}{2} \left( F_{\gamma 2} + G_{\gamma 0} + G_{\gamma 2}\right) P_2\right] , \nonumber \\
\left( \frac{\partial G_\gamma}{\partial \eta}\right)_C &= &   \left(\dot{\kappa} + \dot{\mu}_\gamma \right)\left[ -G_\gamma + \frac{1}{2} \left( F_{\gamma 2} + G_{\gamma 0} + G_{\gamma 2}\right)(1-P_2)\right],
\label{eq:collisionintegralP}
\end{eqnarray}
where we have used the notation of~\cite{MaB},  and the phase space perturbations $F_{\gamma}$ and $G_{\gamma}$ here can be identified with the temperature and polarisation perturbations of equation~(\ref{master}) via $F_{\gamma}  \equiv 4 \Theta$ and $G_{\gamma} \equiv 4 \Theta^{\rm P}$.  

In terms of Legendre moments,  the Boltzmann hierarchies for the photon temperature and polarisation perturbations following from the modified collision integral~(\ref{eq:collisionintegralT})  read respectively~\cite{InteractingDM, Wilkphoton} 
\begin{align}
\begin{split}
 \label{eq:gammahierarchy}
\dot{\delta}_\gamma &= -\frac{4}{3} \theta_\gamma + 4 \dot{\phi}, \\
\dot{\theta}_{\gamma} &= k^{2}\psi + k^{2}\left(\frac{1}{4}\delta_{\gamma} - \sigma_{\gamma}\right)-\dot{\kappa} \left(\theta_{\gamma} - \theta_b\right) -\dot{\mu}_\gamma \left(\theta_{\gamma} - \theta_{\textrm{DM}}\right), \\
\dot{F}_{\gamma 2} & = 2\dot{\sigma}_{\gamma} =  \frac{8}{15}\theta_{\gamma}-\frac{3}{5}kF_{\gamma3}-\frac{9}{5}(\dot{\kappa} + \dot{\mu})\sigma_{\gamma}+\frac{1}{10}(\dot{\kappa} + \dot{\mu})\left(G_{\gamma 0} + G_{\gamma 2}\right), \\
\dot{F}_{\gamma \ell} & = \frac{k}{2 \ell +1} \left[ \ell F_{\gamma (\ell-1)} - (\ell +1) F_{\gamma (\ell+1)} \right] -  (\dot{\kappa}+ \dot{\mu}_\gamma) F_{\gamma \ell}, \qquad \ell \geq 3,
\end{split}
\end{align}
and
\begin{align}
\label{eq:polarisation}
\dot{G}_{\gamma \ell}  = \frac{k}{2 \ell +1} \left[ \ell G_{\gamma (\ell-1)} - (\ell +1) G_{\gamma (\ell+1)} \right]  + (\dot{\kappa}+ \dot{\mu}_\gamma) 
\left[- G_{\gamma \ell} +\frac{1}{2} \Pi \left( \delta_{\ell 0} + \frac{\delta_{\ell 2}}{5}\right)\right],
\end{align}
in the conformal Newtonian gauge, where $\Pi \equiv  F_{\gamma 2} + G_{\gamma 0} + G_{\gamma 2}$, and we identify $\delta_\gamma=F_{\gamma 0}$, \(\theta_{\gamma}=(3/4) k F_{\gamma 1}\) and $\sigma_\gamma= F_{\gamma 2}/2$ as the photon density perturbations, velocity divergence and anisotropic stress respectively.  The corresponding equation of motion for the dark matter perturbations are 
\begin{align} \label{PhotonEuler}
\begin{split}
\dot{\delta}_{\rm DM} & = - \theta_{\rm DM} + 3 \dot{\phi}, \\
\dot{\theta}_{\textrm{DM}} &= k^{2}\psi - \mathcal{H}\theta_{\textrm{DM}} - S_{\gamma}^{-1}\dot{\mu}_\gamma \left(\theta_{\textrm{DM}} - \theta_{\gamma}\right) ,
\end{split}
\end{align}
where $\theta_{\rm DM}= k v_{\rm DM}$,  \(S_{\gamma} \equiv (3/4) \rho_{\textrm{DM}}/\rho_{\gamma}\), and we have again omitted in the Euler equation a pressure gradient term proportional to the square of the DM sound speed.

Note that in the line-of-sight integration approach to computing CMB anisotropies~\cite{Seljak:1996is} used in~\texttt{CLASS}, the visibility and source functions must also be expanded to incorporate the elastic scattering of photons on DM.  We omit writing out these modifications here, since they do not impact on the computation of $\Theta_1$ well before recombination.  The interested reader is referred to~\cite{Wilkphoton} for details.

Because the photon now couples directly to the DM, the modified collision integrals~(\ref{eq:collisionintegralT}) and~(\ref{eq:collisionintegralP}) induce changes in the effective heating rate~(\ref{master})  {\it not only} through the photon transfer functions (such as in the case of DM--neutrino scattering), but also directly on the form of the effective heating rate itself as the new interaction enables the dissipation of perturbations to occur via an additional channel.
We detail in the following how we model the latter effect.


\subsection{Modelling the effective heating rate}
\label{sec:dmgamma}

To compute the effective heating rate in the presence of DM--photon scattering, one in principle needs to expand the photon Boltzmann equation to second order.  However, without doing the full calculation, we deduce based on the form of the collision integral~(\ref{eq:collisionintegralT}) and following the arguments of~\cite{Khatri:2012rt} that the heating rate must have the form
\begin{multline}
\label{BigHeating}
\frac{\textrm{d}\left(Q/\rho_{\gamma}\right)}{\textrm{d}z} = -4  \langle \Theta \frac{\rm d}{{\rm  d} z} \Theta \rangle \\
= \frac{4a\dot{\kappa}}{\mathcal{H}}\int \, \frac{k^{2}\textrm{d}k}{2\pi^{2}}P_{\mathcal{R}}(k)\bigg[\frac{\left(3\Theta_{1}-v_{b}\right)^2}{3}+\frac{9}{2}\Theta_{2}^{2}
- \frac{1}{2}\Theta_{2}\left(\Theta_{0}^{\textrm{P}}+\Theta_{2}^{\textrm{P}}\right) + \sum\limits_{\ell\geq3}(2\ell+1)\Theta_{\ell}^{2} \bigg] \\
+ \frac{4a\dot{\mu}_\gamma}{\mathcal{H}}\int \, \frac{k^{2}\textrm{d}k}{2\pi^{2}}P_{\mathcal{R}}(k)\bigg[
\frac{ \left(3\Theta_{1}-v_{\rm DM}\right)^2}{3}  +\frac{9}{2}\Theta_{2}^{2}
- \frac{1}{2}\Theta_{2}\left(\Theta_{0}^{\textrm{P}}+\Theta_{2}^{\textrm{P}}\right) + \sum\limits_{\ell\geq3}(2\ell+1)\Theta_{\ell}^{2} \bigg] ,
\end{multline}
where the second integral due to DM--photon coupling is formally identical in structure to the first  because of our assumption  that the \(\rm{DM}\)--photon coupling is exactly analogous to the standard photon--baryon coupling.   

As in section \ref{blah}, we assume the baryons and photons to be tightly coupled, which implies \(v_{b} \simeq 3\Theta_{1}\), and the higher-order multipole moments \(\Theta_{\ell \geq 2}\) become progressively smaller with~\(\ell\). Then, setting formally all \(\ell > 2\) terms to zero, equations~(\ref{eq:gammahierarchy}) and~(\ref{eq:polarisation}) can be solved to give 
 the approximate relations \(\Theta_{2}^{\rm{P}} + \Theta_{0}^{\rm{P}} \simeq (3/2)\Theta_{2}\) and \((\dot{\kappa} + \dot{\mu}_{\gamma})\Theta_{2} \simeq (8/15)k\Theta_{1}\), from which we obtain
\begin{equation} \label{HeatingTerms}
\frac{\textrm{d}\left(Q/\rho_{\gamma}\right)}{\textrm{d}z} 
\simeq \frac{4a}{\mathcal{H}}\int \, \frac{k^{2}\textrm{d}k}{2\pi^{2}}\, P_{\mathcal{R}}(k) \, k^2 \, \bigg[\frac{1}{\dot{\kappa} + \dot{\mu}_{\gamma}}\frac{16}{15}\Theta_{1}^{2} 
+\frac{\dot{\mu}_{\gamma}}{3 k^2}  \left(3\Theta_{1}-v_{\rm DM}\right)^2 \bigg]
\end{equation}
for the heating rate~(\ref{BigHeating}).
Comparing this with the standard simplified rate~(\ref{simp}), we see that in the tightly-coupled limit the effects of DM--photon interaction are now condensed into (a)~a modified interaction rate $\dot{\kappa} \to \dot{\kappa} + \dot{\mu}_\gamma$ in the viscosity term of the integrand, and (b)~an additional contribution dependent on the relative DM--photon velocity  (``slippage'')  accounting for heat conduction~\cite{Weinberg:1972kfs}.

To deal with the heat conduction term, we first examine the behaviour of the DM velocity perturbation \(v_{\rm{DM}}\) 
in two regimes of the DM--photon coupling.   In analogy with equation~(\ref{nucross}), we define a dimensionless parameter 
\begin{equation} \label{gammacross}
u_{\gamma} = \frac{\sigma_{\rm{DM}-\gamma}}{\sigma_{\rm T}}\left(\frac{100 \, \rm{GeV}}{m_{\rm{DM}}}\right)
\end{equation}
to quantify the DM--photon scattering rate, and  \(u_\gamma = {u^{0}_\gamma} a^{-2}\), where \(u^{0}_\gamma\) is the present-day value, in the case where the cross section scales with $T^2$.

\paragraph{Strongly-coupled DM} This regime is defined by the condition $S_\gamma^{-1}\dot{\mu}_\gamma \gg k,{\cal H}$, where $S_\gamma^{-1}\dot{\mu}_\gamma$ represents the interaction rate per dark matter particle.  Analogously to baryon perturbations, the dark matter perturbation equations of motion~(\ref{PhotonEuler}) are solved in this limit by~\cite{HuSag1,Hu:1996mn}
\begin{equation}
\begin{aligned}
\label{eq:strongDMsolution}
\theta_{\rm DM}  & \simeq \theta_\gamma - \frac{1}{S_\gamma^{-1}\dot{\mu}_\gamma} \left[ {\cal H}  \theta_\gamma+ \dot{\theta}_{\rm DM} - k^2 \psi  \right]\\
& \simeq \theta_\gamma \mp \frac{{\rm i} \omega}{S_\gamma^{-1}\dot{\mu}_\gamma} \theta_\gamma-  \frac{\omega^2}{S_\gamma^{-2}\dot{\mu}_\gamma^2} \theta_\gamma + \cdots,
\end{aligned}
\end{equation}
where the second approximate equality holds in the $k \gg {\cal H}$ limit when the potential~$\psi$ has decayed away, and we have assumed the solution $\theta_\gamma \sim \exp \left(\pm {\rm i} \int {\rm d}  \eta \; \omega \right)$, with $\omega \simeq k c_s$.%
\footnote{Note that all perturbative quantities $(\delta_{\rm DM}, \theta_{\rm DM}, \delta_\gamma$, etc.) have been defined to be real; we use complex notation here on the understanding that only the real component of an expression is retained.}
  Thus, the DM and photons are very nearly comoving, and upon averaging 
the slippage term in equation~(\ref{HeatingTerms}) evaluates  to
\begin{equation}
\label{eq:approxstrong}
\langle (3 \Theta_1 - v_{\rm DM}) ^2 \rangle \simeq 9  \left(\frac{k}{\sqrt{3} S_\gamma^{-1}\dot{\mu}_\gamma} \right)^2 \langle \Theta_1^2 \rangle
\end{equation}
to leading order in $k /S_\gamma^{-1}\dot{\mu}_\gamma$.

\paragraph{Weakly-coupled DM}  Defined by the condition $\mathcal{H} \ll S_\gamma^{-1} \dot{\mu}_\gamma \ll k$~\cite{InteractingDM}, 
scattering in the weakly-coupled DM regime is not sufficiently rapid to keep $\theta_{\rm{DM}}$ exactly on track with~$\theta_{\gamma}$. Nonetheless, we still expect the  DM perturbations to be ``driven'' by the photon perturbations through the coupling (albeit with $|\theta_{\rm DM}| \ll |\theta_\gamma|$ because of the DM's inertia).
 This implies the DM perturbations oscillate with the same frequency as the photon perturbations so that $\dot{\theta}_{\rm DM} \simeq \pm {\rm i} \omega \, \theta_{\rm DM}\simeq
 \pm {\rm i} k c_s \, \theta_{\rm DM}$, and 
 equation~(\ref{PhotonEuler}) can be solved in this limit by
\begin{equation} 
\begin{aligned}
\label{weakcoupling}
\theta_{\rm DM} & \simeq   \pm \frac{{\rm i} S_\gamma^{-1}  \dot{\mu}_\gamma}{\omega} \left(  \theta_{\rm DM} - \theta_\gamma \right) \\
& \simeq \mp \frac{{\rm i} S_\gamma^{-1} \dot{\mu}_\gamma}{\omega} \theta_\gamma + \frac{S_\gamma^{-2} \dot{\mu}_\gamma^2}{\omega^2} \theta_\gamma + \cdots ,
\end{aligned}
\end{equation}
where we have again ignored the potential and ${\cal H} \theta_{\rm DM}$ terms.
Upon averaging we find for the slippage term
\begin{equation}
\label{eq:approxweak}
\langle (3 \Theta_1 - v_{\rm DM})^2 \rangle \simeq 9 \left[1- \left(\frac{\sqrt{3} S_\gamma^{-1}\dot{\mu}_\gamma}{k}  \right)^2 \right] \langle \Theta_1^2 \rangle
\end{equation}
to next-to-leading order in $S_\gamma^{-1}\dot{\mu}_\gamma/k$, where we have assumed $\langle {\rm Re}({\rm e}^{\pm {\rm i} \int {\rm d}\eta \, \omega}) {\rm Im} ({\rm e}^{\pm {\rm i} \int {\rm d}\eta \, \omega}) \rangle \simeq 0$.


\subsubsection{Interpolation function}
\label{sec:interpolation}

\begin{figure}[t]
\begin{center}
        \includegraphics[width=0.8\textwidth]{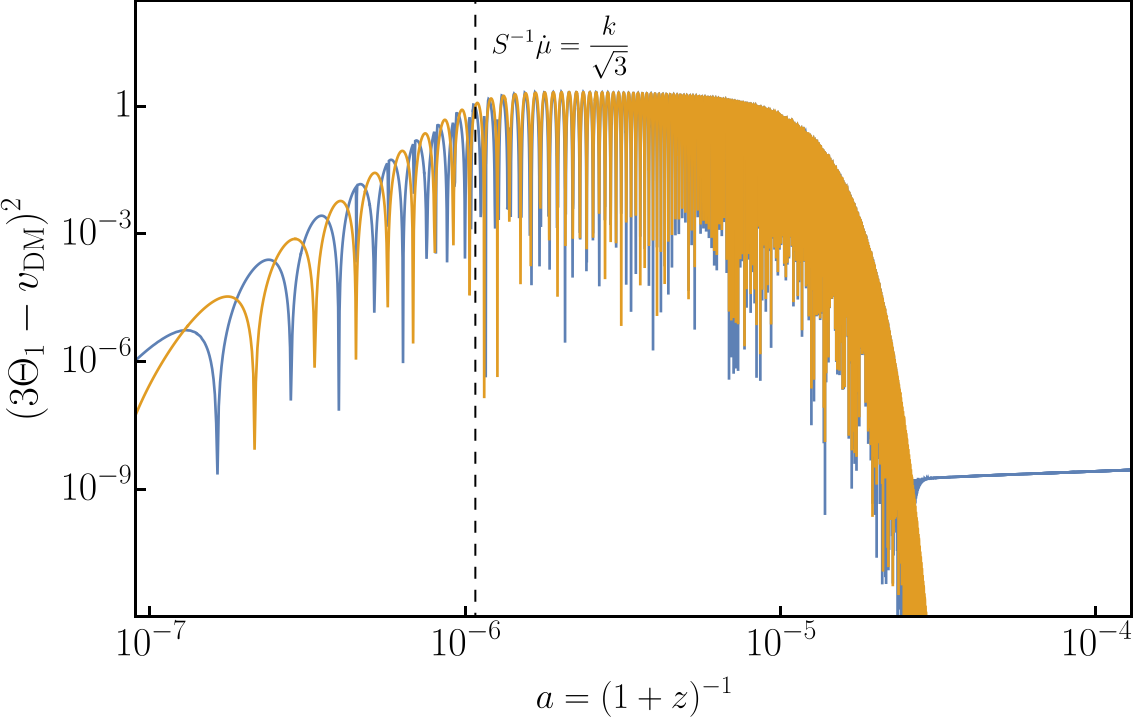}
    \caption{The exact term  \(\left(3\Theta_{1} - v_{\textrm{DM}}\right)^{2}\)  (blue) at \(k = 100 \, \textrm{Mpc}^{-1}\) for a time-independent DM--photon cross section \(u_{\gamma} = 10^{-5}\),
   and its approximation given by equation~(\ref{eq:padeinterpolate})   (orange). 
   The dashed vertical line indicates the transition from strongly-coupled to weakly-coupled DM.}
       \label{TermComp}
    \end{center}
\end{figure}

To connect the strongly-coupled and weakly-coupled DM regimes, we construct an interpolation function in $x\equiv k/(\sqrt{3}S^{-1}_\gamma \dot{\mu}_\gamma)$ by matching the coefficients of a
Pad\'{e} approximant, $R(x) = (a_0 + a_1 x + a_2 x^2 + \cdots)/(1+b_1 x + b_2 x^2+\cdots)$,  
 to the solutions~(\ref{eq:approxstrong}) and (\ref{eq:approxweak})  in the $x\to 0$ and the $x\to \infty$ limit respectively~\cite{Leung:2000kg,Leung:2002bt}.  This procedure yields
\begin{equation}
\label{eq:padeinterpolate}
\langle (3 \Theta_1 - v_{\rm DM})^2 \rangle \simeq 9
 \left(\frac{k^2}{k^2 +3  S^{-2}_\gamma \dot{\mu}_\gamma^2} \right)\
\langle \Theta_1^2 \rangle,
\end{equation}
which, as shown figure~\ref{TermComp} for  $k = 100\, {\rm Mpc}^{-1}$ and $u_\gamma=10^{-5}$, provides a good   approximation to the numerical output of~\texttt{CLASS} in the transition region.
Thus, the heating rate~(\ref{HeatingTerms}) can now be written as
\begin{equation} \label{PhotonHeating}
\frac{\textrm{d}\left(Q/\rho_{\gamma}\right)}{\textrm{d}z} 
\simeq \frac{4a}{\mathcal{H}}\int \, \frac{k^{2}\textrm{d}k}{2\pi^{2}}P_{\mathcal{R}}(k)\, k^2\, \Theta_{1}^{2} \, \left[\frac{1}{\dot{\kappa} + \dot{\mu}_{\gamma}}\frac{16}{15}
+ \frac{3 \dot{\mu}_{\gamma}}{k^2} \left(\frac{k^2}{k^2 +3  S^{-2}_\gamma \dot{\mu}_\gamma^2} \right)\right],
\end{equation}
and we shall be working with this expression in the rest of the analysis.

Observe that while the viscosity term in the integrand~(\ref{PhotonHeating}) is generally a monotonically increasing function of the scale factor $a$, the time dependence of the heat conduction term for $\sigma_{{\rm DM}-\gamma} \propto T^n$ (modulo $
\Theta_1^2$) boils down to
\begin{equation}
\label{eq:burst}
\frac{a \dot{\mu}_\gamma}{\mathcal{H}}  \left(\frac{k^2}{k^2 +3  S^{-2}_\gamma \dot{\mu}_\gamma^2} \right) \, {\rm d} z = {\rm const.} \times 
\frac{(a_{k,{\rm DM}}/a)^n}{1+(a_{k,{\rm DM}}/a)^{2(n+3)}} \, {\rm d} z,
\end{equation}
which is sharply peaked at $a \sim a_{k,{\rm DM}}$, where $a_{k,{\rm DM}}$ is defined via $k = \sqrt{3} S^{-1}_\gamma(a_{k,{\rm DM}}) \dot{\mu}_\gamma(a_{k,{\rm DM}})$.  Thus, we expect DM--photon heat conduction to contribute a temporally localised burst of energy (rather than a sustained injection over time) just as the DM and photon fluids begin to slip past each other.  The degree of localisation depends on the time dependence of the scattering cross section: the larger the $n$ index, the more sharply peaked the burst.

Lastly, we remark  that the DM decouples from the photons when the condition $S_\gamma^{-1}\dot{\mu}_\gamma \ll \mathcal{H}$ is met (i.e., kinetic decoupling). Indeed, we see in figure~\ref{TermComp} that interpolation function~(\ref{eq:padeinterpolate}) eventually ceases to be a faithful approximation to the output of \texttt{CLASS}.  However, in terms of implementation in \texttt{CosmoTherm}, we find that the burst-like time dependence of the heat conduction term ensures that numerically equation~(\ref{PhotonHeating}) suffices to describe the heating rate down to and beyond DM decoupling.  We therefore do not model this transition.


\subsection{Effects on the photon transfer functions}
\label{PhotonEffects}

\begin{figure}[t]
\begin{center}
    \includegraphics[width=0.8\textwidth]{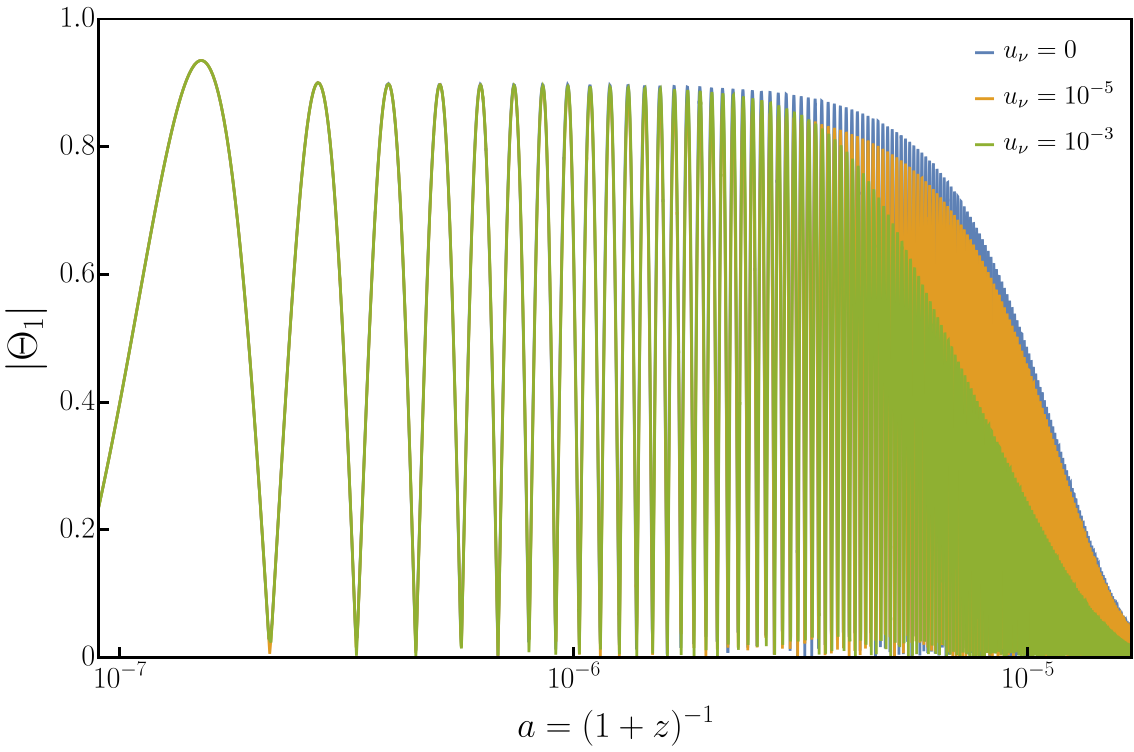}
    \caption{Absolute value of the  photon temperature transfer function \(\Theta_{1}\)  at $k= 100\ {\rm Mpc}^{-1}$ as a function of the scale factor~\(a\) for a selection of time-independent DM--photon scattering cross sections: \(u_{\gamma} = \{10^{-3}, 10^{-5},0\} \), where \(u_{\gamma}\) is defined in equation~(\ref{gammacross}).}
       \label{PhotonFunctions}
    \end{center}
\end{figure}

As in the case of DM--neutrino scattering, DM--photon interaction impacts nontrivially on the photon transfer functions, altering through which the effective heating rate~(\ref{BigHeating}).
Figure~\ref{PhotonFunctions} shows the photon transfer function $\Theta_1$ at  \(k = 100 \, \rm{Mpc}^{-1}\) as a function of the scale factor~\(a\)  for a selection of time-independent DM--photon scattering cross sections.

\begin{figure}[t]
\begin{center}
    \includegraphics[width=0.8\textwidth]{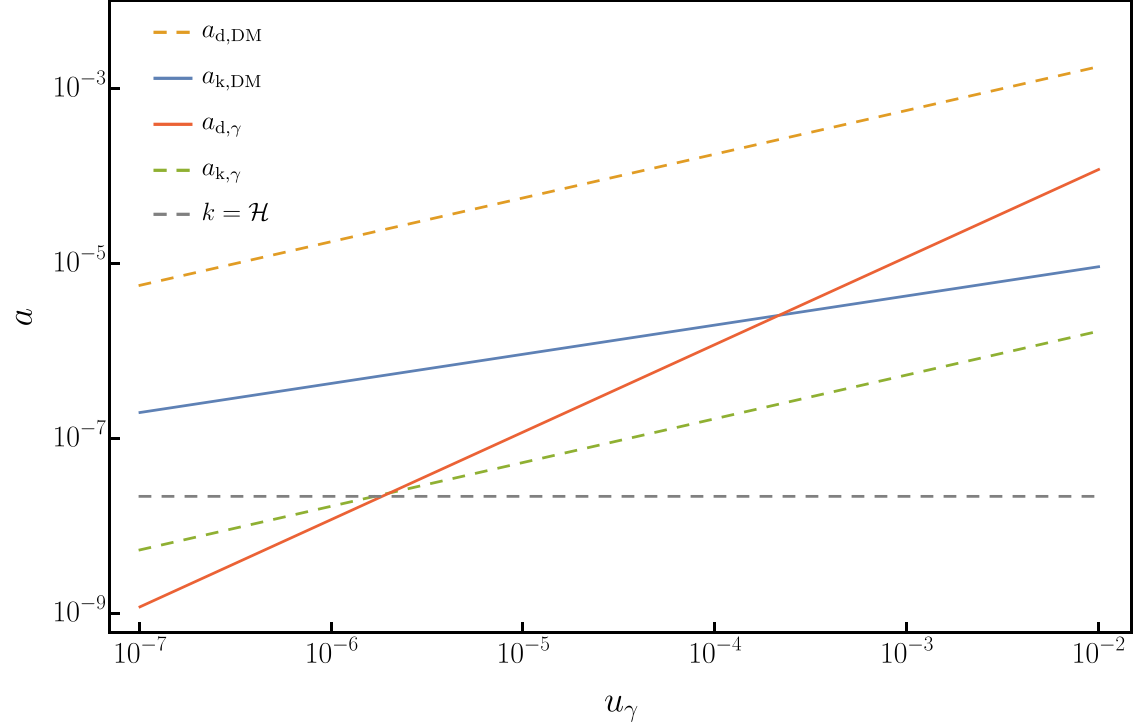}
        \includegraphics[width=0.8\textwidth]{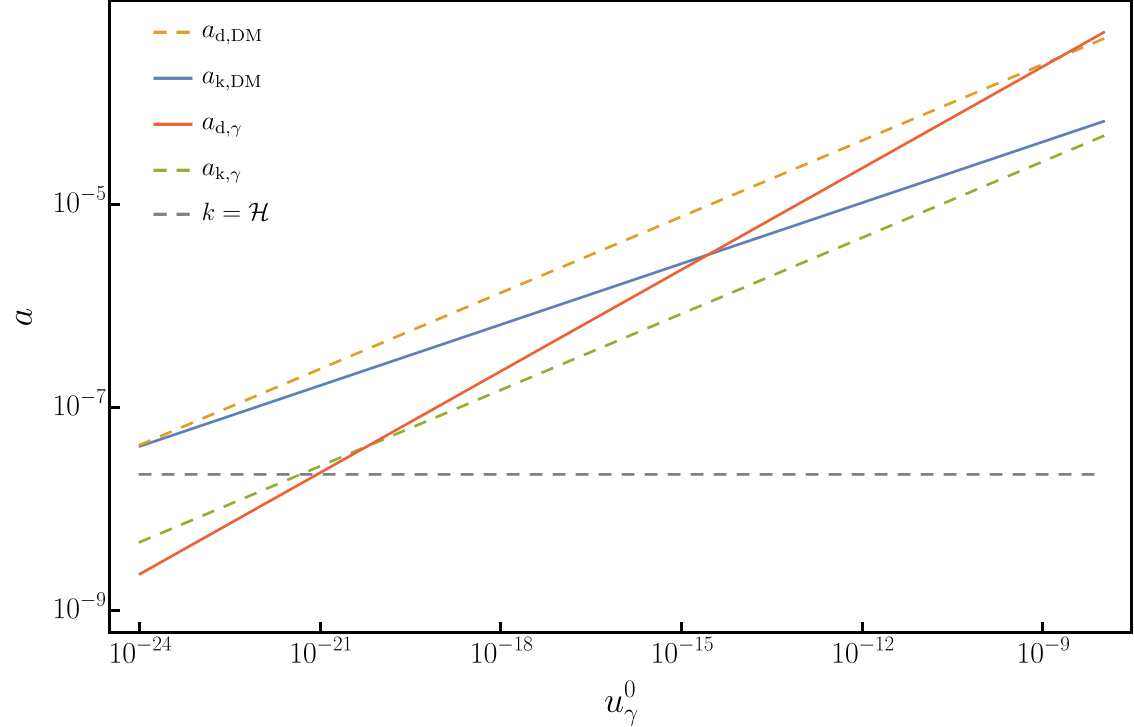}
    \caption{{\it Top}: The four transition epochs---$a_{\rm d,DM}$ (DM kinetic decoupling; orange), $a_{k,{\rm DM}}$ (strongly-coupled to weakly-coupled DM; blue), $a_{{\rm d},\gamma}$ (photon decoupling from DM; red), and $a_{k,\gamma}$ (strongly-coupled to weakly-coupled photons; green)---defined at the beginning of section~\ref{PhotonEffects}, 
 for $k= 100 \ {\rm Mpc}^{-1}$ as functions of a time-independent DM--photon scattering cross section parameterised by $u_\gamma$.        The horizontal grey line indicates the epoch of horizon crossing, i.e., $k  = {\cal H}$.  
 {\it Bottom}: Same as the top panel, but for a scattering cross section scaling as temperature squared.
   \label{fig:a}}
    \end{center}
\end{figure}

To understand the behaviours of $\Theta_1$ in the presence of DM--photon scattering, it is useful to first define the transition epochs between different DM--photon coupling regimes:
\begin{itemize}
\item Strongly-coupled to weakly-coupled photons, defined via $\sqrt{3} \dot{\mu}_\gamma(a_{k,\gamma}) = k(a_{k,\gamma})$, 
\item Weakly-coupled to decoupled photons (from DM), $\dot{\mu}_\gamma(a_{{\rm d},\gamma}) = {\cal H}(a_{{\rm d},\gamma})$, 
\item Strongly-coupled to weakly-coupled DM,  $\sqrt{3} S_\gamma^{-1} \dot{\mu}_\gamma(a_{k,{\rm DM}}) = k(a_{k,{\rm DM}})$, and
\item Weakly-coupled to decoupled DM (i.e., kinetic decoupling),  $S_\gamma^{-1} \dot{\mu}_\gamma(a_{\rm d,DM}) = {\cal H}(a_{\rm d,DM})$.
\end{itemize}
 Figure~\ref{fig:a} shows the four quantities $\{a_{{\rm d},\gamma},a_{k,\gamma}, a_{\rm d,DM},a_{k,{\rm DM}}\}$ as functions of the DM--photon scattering cross section  for $k = 100\ {\rm Mpc}^{-1}$; approximate analytical forms can be found in appendix~\ref{sec:appendix}.  
 Typically, $a_{k,\gamma} > a_{{\rm d},\gamma}, a_{k,{\rm DM}} > a_{\rm d,DM}$ on subhorizon scales, where, as we shall see in the following, it is $a_{k,{\rm DM}}$ and its relation to $a_{{\rm d},\gamma}$ that regulate the effect of DM--photon scattering on the photon transfer functions.

\paragraph{Strongly-coupled photons}
 At very early times $a \ll a_{k,\gamma}$, the condition of strongly-coupled photons generally  has no formal impact on the evolution of the photon perturbations.  This is because Thomson scattering of photons on electrons alone already drives $\Theta_{\ell \geq 2}$ to zero in the tightly-coupled photon--baryon fluid, ensuring that $\Theta_0, \Theta_1$  undergo acoustic oscillations in the spectral distortion timeframe, and solutions of the form $\Theta_1 \simeq A c_s \sin(k r_s) {\rm e}^{-k^2/k_D^2}$ always apply.
 
 Within the damped harmonic oscillator description, the primary role of DM--photon scattering in this regime is to modify the diffusion scale to  $\partial_z k_D^{-2}  \simeq   - (16/15) c_s^2 a/[2 {\cal H} (\dot{\kappa}+ \dot{\mu}_\gamma)]$ through enhanced viscous damping.  
 Furthermore, strongly-coupled photons generally automatically guarantee strongly-coupled DM (i.e., $a_{k,\gamma} < a_{k,{\rm DM}}$), so that the  DM perturbations satisfy \(\theta_{\rm{DM}} \simeq \theta_{\gamma} \).  This leads in principle to a correction to the effective sound speed from DM loading,  $c_s  \equiv  1/\sqrt{3 (1+ R + S_\gamma)} \simeq1/\sqrt{3}$, although in practice the correction is negligibly small.

 \paragraph{Strongly-coupled DM and weakly-coupled or decoupled photons} In the intermediate regime $a_{k,{\rm DM}}, a_{{\rm d},\gamma} \lesssim  a \lesssim a_{k,\gamma}$, where the photons are either only weakly coupled or completely decoupled from the DM while the DM is still strongly coupled to the photons, slippage between the DM and photons begins to creep in and heat conduction becomes possible.
 Here, using the solution~(\ref{eq:strongDMsolution}) for $\theta_{\rm DM}$ up to second order in $\omega/S_\gamma^{-1} \dot{\mu}_\gamma$, the equation of motion for $\theta_\gamma$~(\ref{eq:gammahierarchy}) can be solved in the manner of~\cite{Hu:1996mn} to give a corrected diffusion scale that accounts for this effect:
  \begin{equation}
 \label{eq:diffusionstrong}
\partial_z k_D^{-2} 
 \simeq   -\frac{a}{6 {\cal H} } \left[\frac{1}{\dot{\kappa}+\dot{\mu}_\gamma}\frac{16}{15}+ \frac{3 \dot{\mu}_\gamma }{k^2} \left( \frac{k}{\sqrt{3} S_\gamma^{-1} \dot{\mu}_\gamma}\right)^2  \right].
   \end{equation} 
Note here that the new term (proportional to \(S_{\gamma}^{2}\)) arising from  DM--photon conduction is completely analogous to the  $\propto R^2$ term in equation~(\ref{diffusiondamping}) due to photon--baryon heat conduction and which we have omitted  in the above expression.  We have likewise neglected corrections to the sound speed from baryon and DM loading.

\paragraph{Weakly-coupled DM and weakly-coupled or decoupled photons} As the system moves forward in time to $a \gtrsim a_{k, {\rm DM}}$, the DM perturbations transit to the weakly-coupled DM regime and eventually to kinetic decoupling.  Here,  $\theta_{\rm DM}$ is solved by the approximate solution~(\ref{weakcoupling}), and quickly decays away relative to $\theta_\gamma$.  

For the photon perturbations, this is the regime in which damping via heat conduction with the DM can become very efficient. Using the solution~(\ref{weakcoupling}) for $\theta_{\rm DM}$ up to second order in $S_\gamma^{-1} \dot{\mu}_\gamma/\omega$, we find following the procedure of~\cite{Hu:1996mn} 
 \begin{equation}
 \label{eq:weakdiffusion}
\partial_z k_D^{-2}(k) \simeq -\frac{a}{6 {\cal H} } \left[\frac{1}{\dot{\kappa}+\dot{\mu}_\gamma}\frac{16}{15} + \frac{3 \dot{\mu}_\gamma }{k^2} \left( 1- \left[\frac{\sqrt{3} S^{-1}_\gamma \dot{\mu}_\gamma}{k}  \right]^2 \right) \right] 
\end{equation}
for the diffusion scale.  Observe that in this regime diffusion due to heat conduction is no longer suppressed by $S_\gamma^2$, and can in fact even dominate over viscous damping if $3 \dot{\mu}_\gamma /k  \gtrsim k/\dot{\kappa}$.
This accounts for the early suppression of $\Theta_1$ seen in figure~\ref{PhotonFunctions}, especially in the $u_\nu=10^{-3}$ case, and constitutes the primary signature of DM--photon scattering for  CMB temperature anisotropy measurements~(e.g., \cite{Wilkphoton}).
 Note also that $k_{\rm D}$ is now a $k$-dependent quantity.

 The transition from the weakly-coupled to the decoupled DM regime  at $a \sim a_{\rm d,DM}$ generally has no discernible impact on the photon diffusion scale; by this time viscous damping invariably  dominates over  heat conduction, and equation~(\ref{eq:weakdiffusion}) continues to apply.


\subsubsection{Modelling the photon transfer functions}

To connect equations~(\ref{eq:diffusionstrong}) and~(\ref{eq:weakdiffusion}) through the transition from strongly-coupled to weakly-coupled DM, we appeal again to the Pad\'{e} interpolation function~(\ref{eq:padeinterpolate}) to obtain
 \begin{equation}
 \label{eq:interpolatediffusion}
\partial_z k_D^{-2} (k) \simeq -\frac{a}{6 {\cal H} } \left[\frac{1}{\dot{\kappa}+\dot{\mu}_\gamma}\frac{16}{15} + \frac{3 \dot{\mu}_\gamma }{k^2} \left( 
\frac{k^2}{k^2+ 3 S^{-2}_\gamma \dot{\mu}_\gamma^2}\right) \right] .
\end{equation}
As in the standard case, the diffusion damping scale~$k_D$ sets the damping envelope for $\Theta_1$ in the form $\exp(-k^2/k_D^2)$.
Observe that the envelope from the heat conduction term can be written as
\begin{equation}
\label{eq:dampingenvelope}
\exp\left[-\frac{1}{2} \int_z^\infty  {\rm d}z'\,  \frac{a \dot{\mu}_\gamma}{ {\cal H} }   \left( 
\frac{k^2}{k^2+ 3 S^{-2}_\gamma \dot{\mu}_\gamma^2}\right) \right]= \exp\left[-\frac{1}{2} \left(\frac{a_{{\rm d},\gamma}}{a_{k,{\rm DM}}} \right)^{n+1} \int^\infty_{a_{k,{\rm DM}}/a}  {\rm d} y \,\frac{y^{n}}{1+ y^{2(n+3)}} \right]
\end{equation}
for  $\sigma_{\rm DM-\gamma} \propto T^n$.  Evidently, the integral dictates that heat conduction damping should be switched on at $a_{k,{\rm DM}}/a \simeq 1$.   The damping rate, on the other hand, is regulated by the ratio $a_{{\rm d},\gamma}/a_{k,{\rm DM}}$, which, as shown in figure~\ref{fig:a}, may be $\ll1$ (for very small scattering cross sections) so that the envelope evaluates to essentially unity, or $\gtrsim 1$ (for large cross sections) in which case $\Theta_1$ can become strongly suppressed.
Analytical forms of the integral for $n=0,2$ can be found in appendix~\ref{sec:dampingenvelope}.

Combining the diffusion scale~(\ref{eq:interpolatediffusion}) with the heating rate~(\ref{PhotonHeating}) we obtain
\begin{equation}
\begin{aligned}
\label{eq:finalheatingphoton}
\frac{\textrm{d}\left(Q/\rho_{\gamma}\right)}{\textrm{d}z} 
& \simeq 4 A^2\int \, \frac{k^{2}\textrm{d}k}{2\pi^{2}}P_{\mathcal{R}}(k)\, k^2\,   \frac{a}{6\mathcal{H}}  \left[\frac{1}{\dot{\kappa} + \dot{\mu}_{\gamma}}\frac{16}{15}
+ \frac{3 \dot{\mu}_{\gamma}}{k^2} \left(\frac{k^2}{k^2 +3  S^{-2}_\gamma \dot{\mu}_\gamma^2} \right)\right] \, {\rm e}^{-2 k^2/k_D^2(k)} \\
& =-4 A^2\int \, \frac{k^{2}\textrm{d}k}{2\pi^{2}}P_{\mathcal{R}}(k)\, k^2\,  \left[\partial_z k_D^{-2} (k) \right] \, {\rm e}^{-2 k^2/k_D^2(k)},
\end{aligned}
\end{equation}
where we see that the heating rate essentially reduces to the ``standard form''~(\ref{fullsimplified}) used in~\texttt{CosmoTherm}, with the proviso that $k_D$ must now be modified as per equation~(\ref{eq:interpolatediffusion}).


\subsection{Expected spectral distortions}
\label{Expected}

\begin{figure}[t]
\begin{center}
    \includegraphics[width=0.8\textwidth]{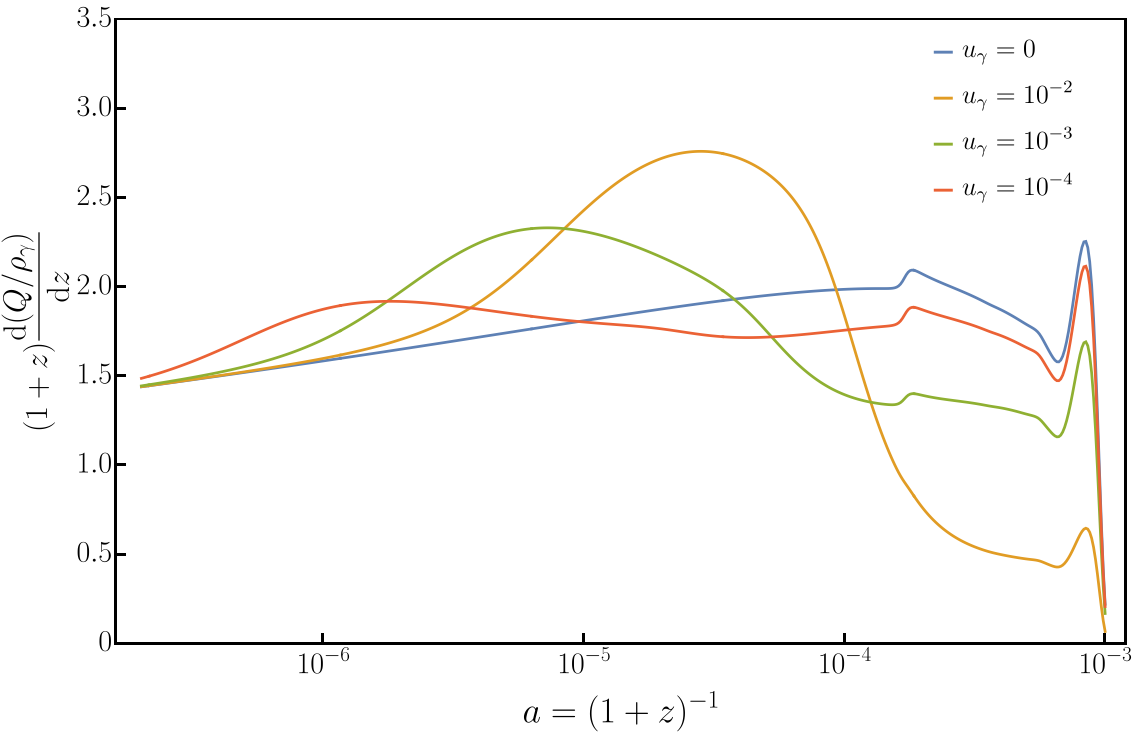}
    \includegraphics[width=0.8\textwidth]{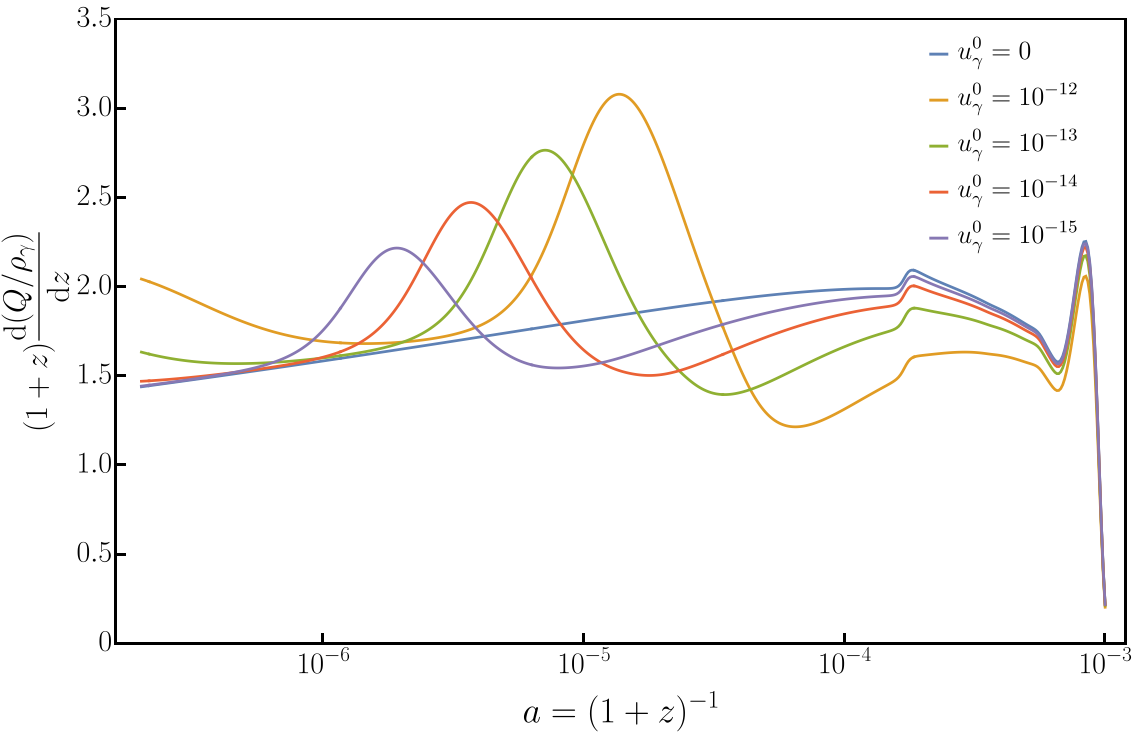}
    \caption{\textit{Top:} Effective heating rate as a function of the scale factor \(a\) for a selection of DM--photon scattering cross sections:  $u_\gamma=\{10^{-2}, 10^{-3}, 10^{-4},0 \}$.  In all cases the amplitude of the primordial power spectrum has been set to unity \(A_s = 1\) at the pivot scale $k_0=0.05\ {\rm Mpc}^{-1}$, the spectral index to $n_s = 0.96$, and we assume no running $n_{\rm run}=0$. All other cosmological parameters assume the Planck 2015 vanilla best-fit values. 
   \textit{Bottom:} Same as the top panel, but for  DM--photon scattering cross sections proportional to the temperature squared: $u^{0}_{\gamma}=\{10^{-12}, 10^{-13}, 10^{-14}, 10^{-15},0\}$.}
       \label{HeatingRate}
    \end{center}
\end{figure}

Figure~\ref{HeatingRate} shows the heating rate for a selection of DM--photon scattering rates computed using a modified version of~\texttt{CosmoTherm} in accordance with  equation~(\ref{eq:finalheatingphoton}). The background cosmology is taken to be standard vanilla $\Lambda$CDM  specified by the best-fit values to the Planck 2015 data.
Three distinctive features can be discerned in the curves.

\paragraph{Enhancement at very early times}  At early times viscous damping dominates over heat conduction, so that $\partial_z k_D^{-2}  \simeq   - (16/15) c_s^2 a/[2 {\cal H} (\dot{\kappa}+ \dot{\mu}_\gamma)]$, which is independent of the wavenumber~$k$.  Furthermore, the conformal Thomson scattering rate scales  as \(\dot{\kappa} \propto a^{-2}\), while the DM--photon scattering rate follows  \(\dot{\mu}_\gamma \propto a^{-(n+2)}\) if the DM--photon scattering cross section  should scale as $T^n$.  This implies that for $n>0$, DM--photon scattering can in fact overtake Thomson scattering as the dominant viscous dissipation channel at very early times. 

Thus, assuming an almost scale-invariant primordial power spectrum, equation~(\ref{eq:finalheatingphoton}) can be integrated to give
\begin{equation}
\label{eq:earlyheating}
\frac{\textrm{d}\left(Q/\rho_{\gamma}\right)}{d z} \simeq -A^{2} \, {k_{D}}^{2}\left(\partial_{z} k_{D}^{-2}\right) \simeq
\begin{cases}
3 A^{2}, & \qquad  \dot{\mu}_\gamma \ll \dot{\kappa} \\
(n+3) A^{2}. & \qquad   \dot{\mu}_\gamma \gg \dot{\kappa}
 \end{cases}
\end{equation}
Evidently,  the heating rate for $n>0$ asymptotes to a higher value at early times, which explains the enhancement seen at $a \lesssim 10^{-6}$ in the $\sigma_{{\rm DM}-\gamma} \propto T^2$ case (bottom panel of figure~\ref{HeatingRate}) but is absent when the scattering cross section is time-independent (top panel).

\paragraph{Bump at intermediate times}  DM--photon heat conduction kicks in when the dark matter transits from being strongly-coupled to weakly-coupled to the photons, causing an enhancement in the heating rate via the second term in equation~(\ref{eq:finalheatingphoton}) relative to the $\Lambda$CDM case.  As shown in equation~(\ref{eq:burst}), the contribution from each $k$-mode to this enhancement is in the form of a temporally localised burst at $a \sim a_{k,{\rm DM}}$, where a scattering cross section that decreases in time ($n>0$) tends to produce a more squeezed burst.  This is further aided by the damping envelope~(\ref{eq:dampingenvelope}), which ensures a more rapid suppression of power in $\Theta_1$ for $n>0$.  Indeed, comparing the cases of a constant cross section and $\sigma_{{\rm DM}-\gamma} \propto T^2$, we see in figure~\ref{HeatingRate} that the former's bump is relatively spread out in time, while the latter's is narrower and more well defined.

\paragraph{Suppression at late times}  As we move forward in time, viscous damping resumes its dominance of the heating rate at late times.  Here, $\dot{\mu}_\gamma \ll \dot{\kappa}$, so that Thomson scattering is the main channel of dissipation. The integrand peaks as usual at \(k = k_{D}/2 \simeq 2\times 10^{-6} \, a^{-3/2}\), and heating proceeds from  large to small $k$-modes.  However, because DM--photon heat conduction has already sapped the larger $k$-modes of energy at intermediate times,  the photon perturbations are by late times suppressed by the damping envelope~(\ref{eq:dampingenvelope}) relative to the $\Lambda$CDM case.  This leads to a suppressed heating rate intermediately after the heat conduction bump.

The heating rate eventually tends towards the $\Lambda$CDM rate at very late times, when the integral becomes dominated by small $k$-modes.  Small $k$-modes do not suffer strongly from  damping by DM--photon heat conduction, because the transition from the strongly-coupled to the weakly-coupled DM regime occurs late relative to photon decoupling from DM, i.e., $a_{k,{\rm DM}} \gg a_{{\rm d},\gamma}$.  Thus, the heat conduction damping rate is strongly suppressed by $(a_{{\rm d},\gamma}/a_{k,{\rm DM}})^{n+1}$ (see figure~\ref{fig:a}), so that  the envelope~(\ref{eq:dampingenvelope}) evaluates essentially to unity, and we recover the $\Lambda$CDM heating rate.

\bigskip

\begin{figure}[t]
\begin{center}
    \includegraphics[width=0.9\textwidth]{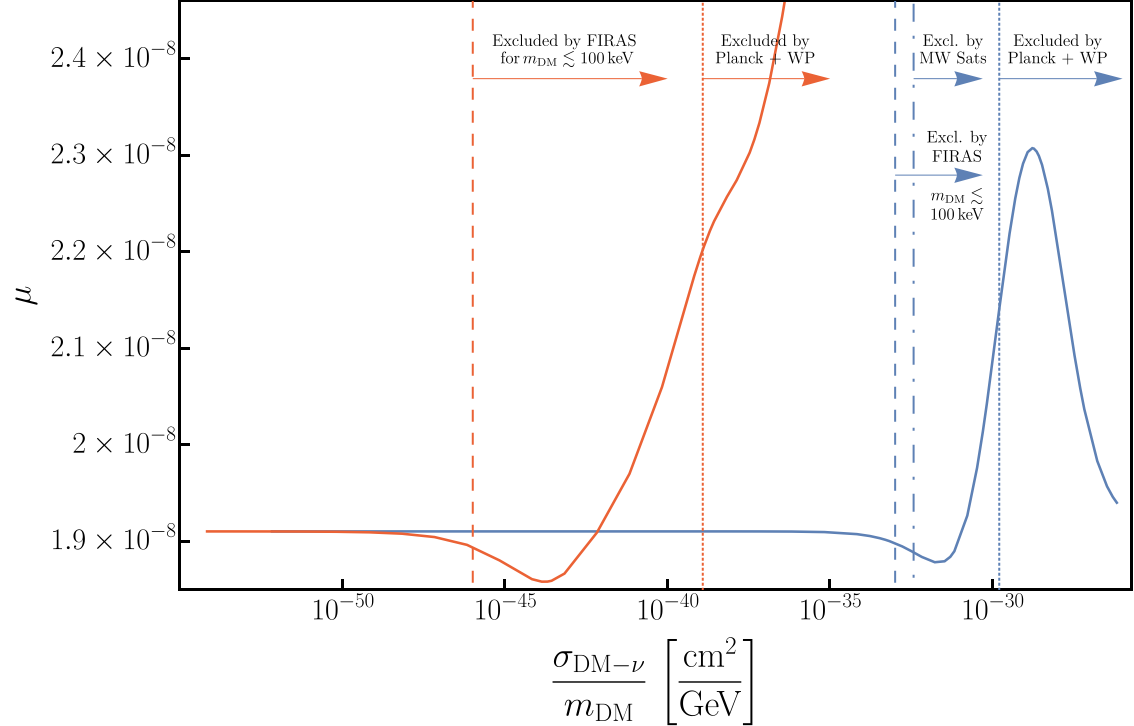}
    \caption{Expected \(\mu\)-parameter as a function of the  present-day DM--photon scattering cross section.  Here, the blue line denotes the case of a time-independent cross section, and the red line the case in which the cross section is proportional to the temperature squared. In both cases we assume a primordial power spectrum amplitude of \(A_{s} = 2.2\times10^{-9}\) at the pivot scale \(k_{0} = 0.05 \, \textrm{Mpc}^{-1}\), a spectral index  \(n_{s} = 0.96\), and no running  \(n_{\textrm{run}} = 0\).  The vertical lines indicate the parameter regions presently excluded by various  observations at 95\%~C.L. }
       \label{PhotonMu}
    \end{center}
\end{figure}

Figure~\ref{PhotonMu} shows the expected $\mu$-distortion as a function of the DM--photon elastic scattering cross section, assuming the cases of a time-independent $\sigma_{\textrm{DM}-\gamma}$ and $\sigma_{\textrm{DM}-\gamma} \propto T^2$.
In contrast to the case of DM--neutrino scattering which always leads to an augmented $\mu$-parameter relative to the base $\Lambda$CDM value, the overall effect of DM--photon scattering on $\mu$ can work in either direction depending on the magnitude and time evolution of the scattering cross section. Indeed, the nontrivial shapes of the  $\mu$-distortion curves can be easily understood
in terms of which of the three features identified above dominates the heating rate during the $\mu$-era ($3\times 10^{5} \lesssim z \lesssim 2 \times10^6$): For large cross sections the $\mu$-era coincides predominantly with the heat conduction bump and/or the early enhancement from increased viscosity (if $\sigma_{\textrm{DM}-\gamma} \propto T^2$), leading to an enhanced $\mu$-parameter relative to $\Lambda$CDM. For small cross sections it is the post-heat conduction suppression that falls in the $\mu$-era; the value of $\mu$ decreases
correspondingly (but remains positive).

\section{Implications for future observations}
\label{sec:PRISM}

Recall that the projected 1\(\sigma\) sensitivities for a PIXIE- and a PRISM-type experiment are \(\sigma(|\mu|) \sim 10^{-8}\) and \(\sigma(|\mu|) \sim 10^{-9}\) respectively~\cite{PIXIE1,PIXIE2,Andre:2013afa}. Given that, relative to $\Lambda$CDM, DM--$X$ elastic scattering typically produces deviations of order $10^{-9}$ in the $\mu$-parameter, there is clearly no prospect for detecting/constraining these scenarios using PIXIE.  However, these deviations may still be within the reach of a PRISM-like experiment.

\begin{figure}[t]
\begin{center}
    \includegraphics[width=0.8\textwidth]{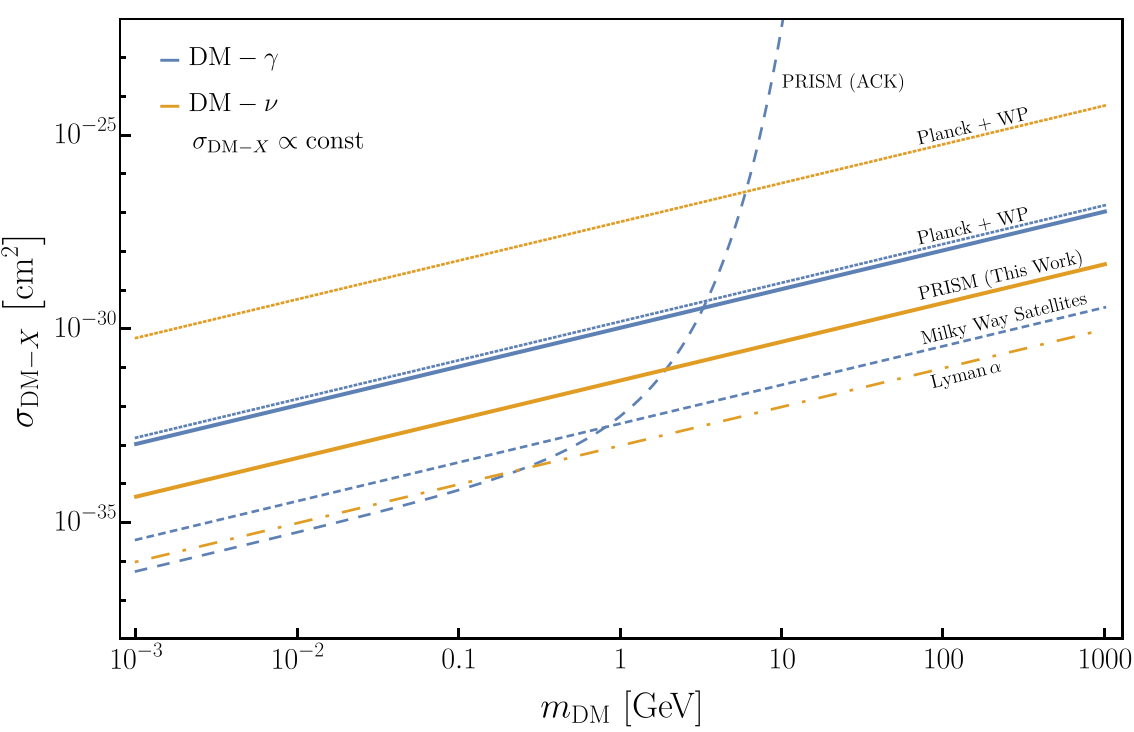}
    \includegraphics[width=0.8\textwidth]{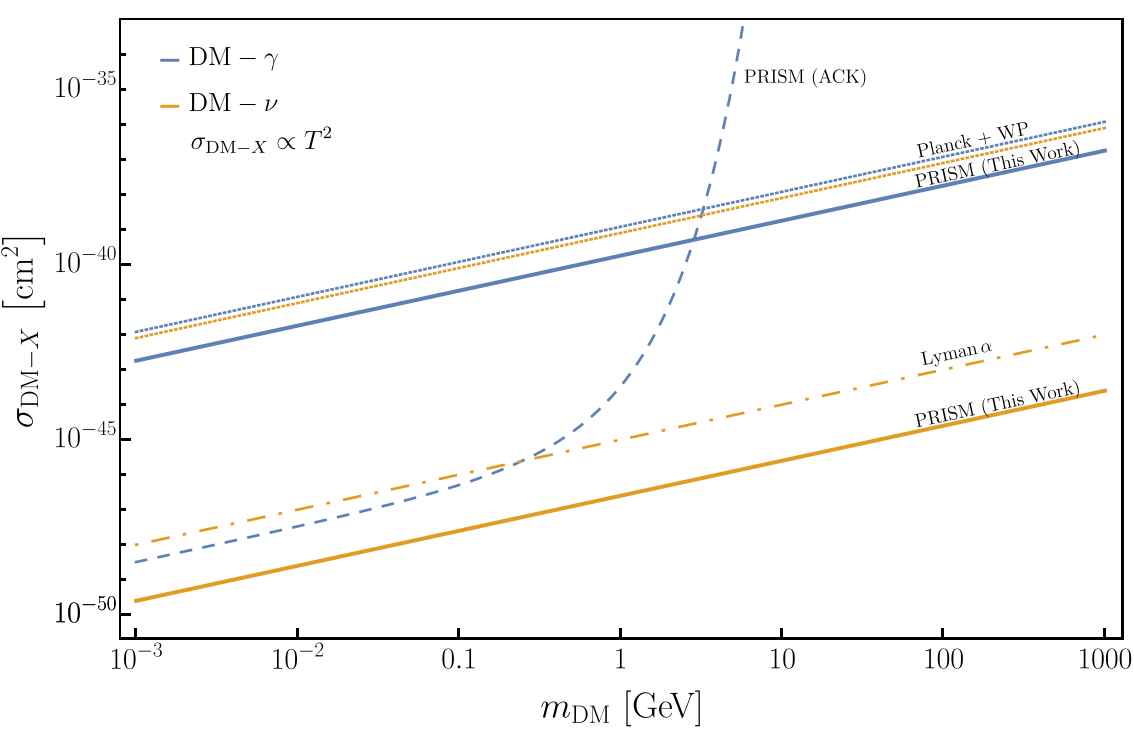}
    \caption{\textit{Top:} Projected sensitivity regions of PRISM to DM--photon (blue) and DM--neutrino (orange) elastic scattering.  Solids lines indicate the \(2 \sigma\) sensitivity reach of a PRISM-like experiment derived in this work, while the long-dashed lines labelled ``PRISM (ACK)'' correspond to projected PRISM constraints on the effect discussed in~\cite{ChlubaKam}.  We also 
  show the 95\% C.L.\ upper limits from Planck+WP (dotted), the Lyman-$\alpha$ forest (dot-dashed), and Milky Way satellite counts (short-dashed) summarised in table~\ref{tab:current constraints}.
    \textit{Bottom:} Same as top panel, but for scattering cross sections proportional to temperature squared.}
       \label{DMphotonbound}
    \end{center}
\end{figure}

In the following, we consider what elastic scattering cross section it takes to produce a ``$2 \sigma$'' deviation of $\Delta \mu = 2 \times 10^{-9}$, assuming that all other cosmological parameters can be constrained (by other means) to such an accuracy so as not to impact on $\mu$ at a similarly significant level.  We emphasise that this is by no means a proper forecast; our purpose is to provide a first assessment of the potential reach of a PRISM-like experiment for LKD scenarios.  Within this interpretation, figure~\ref{DMphotonbound} summarises the potential sensitivity regions of a PRISM-like experiment relative to current constraints.

\paragraph{Dark matter--neutrino scattering} Because the enhancement in $\mu$ is capped at $\sim 20$\% of the base $\Lambda$CDM value (see figure~\ref{NeutrinoMu}), we see that the prospects for detecting DM--neutrino scattering by a PRISM-like experiment are $\sim 4\sigma$ at maximum.  At $2 \sigma$ we find that PRISM will be sensitive to DM--neutrino elastic scattering cross sections in the region
\begin{align}
\begin{split}
\label{eq:sensitivityNu}
\sigma_{\textrm{DM}-\nu} &\gtrsim 4.8\times 10^{-32} \, \left(m_{\textrm{DM}}/\textrm{GeV}\right) \, \textrm{cm}^{2}, \qquad \, \, \sigma_{\textrm{DM}-\nu} = \textrm{const.}  \\
\sigma^{0}_{\textrm{DM}-\nu} &\gtrsim 2.5 \times 10^{-47} \, \left(m_{\textrm{DM}}/\textrm{GeV}\right) \, \textrm{cm}^{2}, \qquad \, \, \sigma_{\textrm{DM}-\nu} \propto T^{2}
\end{split}
\end{align}
improving the sensitivity reach of Planck+WP by about four and eight orders of magnitude respectively.  Note also that future CMB anisotropy measurements are unlikely to improve on current Planck bounds by more than a factor of a few~\cite{Escudero2015}.

Relative to low-redshift large-scale structure observations, we find that $\mu$-distortion with PRISM falls short of current Lyman-$\alpha$ bounds by one to two orders of magnitude in the case of a time-independent cross section, while for $\sigma_{{\rm DM}-\nu} \propto T^2$ it may outperform the Lyman-$\alpha$  sensitivity reach by a similar stretch.  Most importantly, because the parameter regions probed by spectral distortions and by the Lyman-$\alpha$ forest are comparable, a PRISM-like experiment will potentially provide an independent high-redshift verification of the low-redshift exclusion limits on DM--neutrino scattering, the latter of which are arguably more susceptible to modelling uncertainties due to nonlinear evolution.%
 \footnote{We note here that the Lyman-$\alpha$ bound on  $\sigma_{{\rm DM}-\nu}/m_{\rm DM}$ had  {\it not} in fact been derived in~\cite{Wilknu} using a full hydrodynamic simulation of the LKD scenario, but rather from matching the LKD linear power spectrum to that of a thermal WDM with the smallest mass allowed by observations.}
 
\paragraph{Dark matter--photon scattering} At $2 \sigma$ we find that  PRISM is potentially sensitive to $\mu$-distortions due to DM--photon scattering cross sections in the ballpark
 \begin{align}
\begin{split}
\label{eq:sensitivityPhoton}
\sigma_{\textrm{DM}-\gamma} &\gtrsim 1.1\times 10^{-30} \, \left(m_{\textrm{DM}}/\textrm{GeV}\right) \, \textrm{cm}^{2}, \qquad \, \, \sigma_{\textrm{DM}-\gamma} = \textrm{const.}  \\
\sigma^{0}_{\textrm{DM}-\gamma} &\gtrsim 1.8 \times 10^{-40} \, \left(m_{\textrm{DM}}/\textrm{GeV}\right) \, \textrm{cm}^{2}, \qquad \, \, \sigma_{\textrm{DM}-\gamma} \propto T^{2}
\end{split}
\end{align}
providing a modest (up to an order of magnitude) improvement over current CMB anisotropy bounds from Planck+WP, but falling short of Milky Way satellite constraints by more than two orders of magnitude (in the constant $\sigma_{{\rm DM}-\gamma}$ case).  It is also interesting to note that, comparing equations~(\ref{eq:sensitivityNu}) and~(\ref{eq:sensitivityPhoton}), the sensitivity of PRISM to DM--photon scattering is considerably diminished relative to DM--neutrino scattering.  This can be traced to the competing effects of heat conduction and late-time viscous damping (see figure~\ref{HeatingRate}), which cancel each other out when integrated in time over the duration of the $\mu$-era.

As already discussed in section~\ref{sec:microphysics}, the null detection by FIRAS of a negative $\mu$ due to kinetic energy transfer from photons to the DM also places a limit on $\sigma_{{\rm DM}-\gamma}/m_{\rm DM}$ for DM masses  $m_{\rm DM} \lesssim 100$~keV~\cite{ChlubaKam}.  The role of PRISM will be  to extend this constraint up to $m_{\rm DM} \sim 10$~GeV.  As shown  in figure~\ref{DMphotonbound}, this effect---labelled ``PRISM (ACK)''---will be the dominant source of $\mu$-distortion for DM masses up to $\sim 3$~GeV;  Beyond that distortions due to the dissipation of small-scale perturbations take over, thereby extending the sensitivity reach of a PRISM-like experiment to include heavy DM masses. There is also a small region at $m_{\rm DM} \sim 3$~GeV in  which the two effects will cancel because of their opposite signs.

Lastly, we remark that  given the pronounced time dependence of the heating rate demonstrated in~figure~\ref{HeatingRate}, it is possible that  the $r$-distortion (see equation~(\ref{eq:distort})) may provide a more sensitive probe of DM--photon interactions.  We defer the investigation of this possibility to a later work.


\section{Conclusions}
\label{sec:conc}

In this work, we have computed the dissipation of small-scale perturbations in the early universe in two late kinetic decoupling dark matter scenarios, and their subsequent contributions to distortions of the $\mu$-type in the CMB energy spectrum.

For dark matter--neutrino elastic scattering, we find that the amplitude of the photon--baryon acoustic oscillations can be enhanced as a consequence of the reduction of neutrino anisotropic stress at horizon crossing brought on by the interaction.  This is purely a gravitational effect, and causes a maximum $\sim 20$\% increase in the $\mu$-parameter relative to the $\Lambda$CDM prediction, and may be distinguishable from $\Lambda$CDM by a PRISM-like experiment if the present-day value of the scattering cross section is at least as large as 
\(\sigma_{\textrm{DM}-\nu} \gtrsim 4.8 \times 10^{-32}  \left(m_{\textrm{DM}}/\textrm{GeV}\right) \, \textrm{cm}^{2}\) 
for time-independent cross sections, and \(\sigma^{0}_{\textrm{DM}-\nu} \gtrsim 2.5 \times 10^{-47}  \left(m_{\textrm{DM}}/\textrm{GeV}\right) \, \textrm{cm}^{2}\) for $\sigma_{{\rm DM}-\gamma} \propto T^2$.   In the latter case, it is interesting to note that the constraining power of PRISM on dark matter--neutrino elastic scattering may potentially exceed current limits from the Lyman-\(\alpha\) forest.

For dark matter--photon elastic scattering, we find that the effective photon heating rate must be altered in a nontrivial way to account for the fact that dark matter--photon scattering explicitly provides an additional channel through which small-scale fluctuations may dissipate. We derive new analytical expressions for the diffusion scale and the heating rate for a tightly-coupled photon--baryon fluid including a new dark matter--photon coupling term that is gradually switched off. In contrast  to the $\Lambda$CDM  case (and also the case of dark matter--neutrino scattering) wherein  dissipation occurs almost exclusively through shear viscosity in the photon fluid, dark matter--photon scattering enables in addition dissipation through heat condition when the dark matter and photon fluids begin to slip past each other.  Indeed, such heat conduction can even be the dominant mode of dissipation in the $\mu$-era.
 The resulting $\mu$-distortion may be diminished or enhanced relative to the $\Lambda$CDM prediction, and distinguishable from $\Lambda$CDM by a PRISM-like experiment if \(\sigma_{\textrm{DM}-\gamma} \gtrsim 1.1 \times 10^{-30} \left(\textrm{cm}^{2}/\textrm{GeV}\right)\) (constant $\sigma_{{\rm DM}-\gamma}$), or \(\sigma^{0}_{\textrm{DM}-\gamma} \gtrsim 1.8 \times 10^{-40} \left(\textrm{cm}^{2}/\textrm{GeV}\right)\) ($\sigma_{{\rm DM}-\gamma} \propto T^2$).
 
  Previous works have shown that a PRISM-like experiment will be able to put stringent limits on the dark matter--photon elastic scattering cross section for small dark matter masses,  \(m_{\textrm{DM}} \lesssim 1\)~GeV, based on a $\mu$-distortion arising from the transfer of kinetic energy~\cite{ChlubaKam}.  In the present work we have shown that $\mu$-distortions from the dissipation of small-scale perturbations can extend the sensitivity reach of PRISM to this interaction to much heavier dark matter masses.

Allowing the dark matter to decouple kinetically from neutrinos or photons at a relatively late time is a parsimonious and theoretically appealing way to circumvent the small-scale problems faced by conventional cold dark matter cosmology.  Compared with the popular warm dark matter solution which has no $\mu$-distortion signal discernible from the $\Lambda$CDM prediction, late kinetic  decoupling scenarios predict deviations in the $\mu$-parameter that are potentially within the reach of future experiments such as PRISM.  The signals also differ from that of a primordial suppression scenario recently considered in~\cite{Nakama:2017ohe}, wherein the small-scale crisis is solved by a highly suppressed primordial curvature power spectrum rather than through late-time dynamical means; such a scenario predicts a highly suppressed $\mu$-parameter, which may even turn negative for certain model parameters.  

Thus, while these scenarios all have broadly similar low-redshift phenomenologies, the detection of CMB spectral distortions could potentially contribute to lifting the degeneracy through an independent high-redshift verification of the physics on the relevant length scales, and hence provide a handle to distinguishing between solutions to the small-scale crisis.


\acknowledgments
We thank J. Hamann and S. Hannestad for discussions, T.~Tram and R.~Wilkinson for useful hints on the implementation of WDM and LKD in \texttt{CLASS}, and J.~Chluba for clarifications concerning \texttt{CosmoTherm}.
JADD acknowledges support from an Australian Government Research Training Program Scholarship.
 The work of Y$^3$W is supported in part by the Australian Government through the Australian Research Council's
Discovery Projects funding scheme (project DP170102382).


\appendix

\section{Definitions of decoupling epochs for dark matter--photon scattering}
\label{sec:appendix}

We summarise here the definitions of various epochs relevant for DM--photon elastic scattering, and their numerical estimates based on the Planck 2015 vanilla best-fit parameter values.
\begin{itemize}
\item We define the transition from strongly-coupled to weakly-coupled DM via  the relation $\sqrt{3} S_\gamma^{-1} \dot{\mu}_\gamma(a_{k,{\rm DM}}) = k(a_{k,{\rm DM}})$, where $a_{k,{\rm DM}}=a_{k,{\rm DM}}(k,u_\gamma) $ evaluates to
\begin{equation}
\label{eq:akDM}
a_{k,{\rm DM}}  \simeq 
\begin{cases}
 \left(4 \sqrt{3} H_0^2 M_{\rm pl}^2 \Omega_\gamma \frac{1}{k} \frac{\sigma_{{\rm DM}-\gamma}}{m_{\rm DM}} \right)^{1/3} \simeq 0.00020\, u_\gamma^{1/3} \left(\frac{{\rm Mpc}^{-1}}{k}  \right)^{1/3}, & \; \sigma_{\rm DM-\gamma} = {\rm const.} \\
\left(4\sqrt{3}  H_0^2 M_{\rm pl}^2 \Omega_\gamma \frac{1}{k} \frac{\sigma_{{\rm DM}-\gamma}}{m_{\rm DM}} \right)^{1/5} \simeq 0.0066 \, u_\gamma^{1/5} \left(\frac{{\rm Mpc}^{-1}}{k}  \right)^{1/5}, & \; \sigma_{\rm DM-\gamma}  \propto T^2 
 \end{cases}.
\end{equation}

\item The transition from weakly-coupled to decoupled DM, i.e., DM kinetic decoupling, is specified by  $S_\gamma^{-1} \dot{\mu}_\gamma(a_{\rm d,DM}) = {\cal H}(a_{\rm d,DM})$,
where $a_{\rm d,DM}=a_{\rm d,DM}(u_\gamma)$ is given by
\begin{equation}
\label{eq:adDM}
a_{\rm d,DM}(u_\gamma)  \simeq 
\begin{cases}
\left(4 H_0 M_{\rm pl}^2 \frac{\Omega_\gamma}{\sqrt{\Omega_r}} \frac{\sigma_{{\rm DM}-\gamma}}{m_{\rm DM}} \right)^{1/2} \simeq 0.0018  \, u_\gamma^{1/2}, &\quad \sigma_{\rm DM-\gamma} = {\rm constant} \\
\left(4 H_0 M_{\rm pl}^2 \frac{\Omega_\gamma}{\sqrt{\Omega_r}} \frac{\sigma^0_{{\rm DM}-\gamma}}{m_{\rm DM}} \right)^{1/4} \simeq  0.043 \, {u^0_\gamma}^{1/4}, &\quad \sigma_{\rm DM-\gamma} \propto T^2
\end{cases}.
\end{equation}

\item The relation $ \sqrt{3} \dot{\mu}_\gamma(a_{{\rm d},\gamma}) = k(a_{{\rm d},\gamma})$ defines the transition from a regime in which the photons are strongly coupled to DM to one in which their coupling to DM is weak.  Here, $a_{k,\gamma} =a_{k,\gamma}(k,u_\gamma)$, and
\begin{equation}
\label{eq:akgamma}
a_{k,\gamma} \simeq  
\begin{cases}
 \left(3 \sqrt{3}  H_0^2 M_{\rm pl}^2 \Omega_{\rm DM} \frac{1}{k} \frac{\sigma_{{\rm DM}-\gamma}}{m_{\rm DM}} \right)^{1/2} \simeq 0.00017 \, u_\gamma^{1/2} \left(\frac{{\rm Mpc}^{-1}}{k} \right)^{1/2}, &\; \sigma_{\rm DM-\gamma} = {\rm const.} \\
  \left(3 \sqrt{3} H_0^2 M_{\rm pl}^2 \Omega_{\rm DM} \frac{1}{k} \frac{\sigma_{{\rm DM}-\gamma}}{m_{\rm DM}} \right)^{1/4} \simeq  0.015 \, {u^0_\gamma}^{1/4} \left(\frac{{\rm Mpc}^{-1}}{k} \right)^{1/4}, &\; \sigma_{\rm DM-\gamma} \propto T^2
  \end{cases}.
\end{equation}

\item Photons decouple from DM when $\dot{\mu}_\gamma(a_{k,\gamma}) = {\cal H}(a_{k,\gamma})$ is satisfied, where $a_{{\rm d},\gamma}=a_{{\rm d},\gamma}(u_\gamma)$ is given by
\begin{equation}
\label{eq:adgamma}
a_{{\rm d},\gamma}(u_\gamma) \simeq
\begin{cases}
3 H_0 M_{\rm pl}^2 \frac{\Omega_{\rm DM}}{\sqrt{\Omega_r}} \frac{\sigma_{{\rm DM}-\gamma}}{m_{\rm DM}}  \simeq 0.012  \, u_\gamma,  &\quad \sigma_{\rm DM-\gamma} = {\rm const.} \\ 
\left(3 H_0 M_{\rm pl}^2 \frac{\Omega_{\rm DM}}{\sqrt{\Omega_r}} \frac{\sigma_{{\rm DM}-\gamma}}{m_{\rm DM}}  \right)^{1/3} \simeq 0.23  \, {u_\gamma^0}^{1/3},  &\quad \sigma_{\rm DM-\gamma} \propto T^2
\end{cases}.
\end{equation}
\end{itemize}


\section{Damping envelope due to dark matter--photon heat conduction}
\label{sec:dampingenvelope}

Equation~(\ref{eq:dampingenvelope}) gives the damping envelope for the photon transfer function~$\Theta_1$ due to heat conduction between photons and the dark matter from elastic scattering.  The expression is reproduced here:
\begin{equation}
\exp\left[-\frac{1}{2} \int_z^\infty  {\rm d}z'\,  \frac{a \dot{\mu}_\gamma}{ {\cal H} }   \left( 
\frac{k^2}{k^2+ 3 S^{-2}_\gamma \dot{\mu}_\gamma^2}\right) \right]= \exp\left[-\frac{1}{2} \left(\frac{a_{{\rm d},\gamma}}{a_{k,{\rm DM}}} \right)^{n+1} \int^\infty_{a_{k,{\rm DM}}/a}  {\rm d} y \,\frac{y^{n}}{1+ y^{2(n+3)}} \right],
\end{equation}
where the index $n$ denotes the dependence of the DM--photon scattering cross section on the temperature as $\sigma_{{\rm DM}-\gamma} \propto T^n$.

For $n=0, 2$, we find that the $y$-integral evaluates to
\begin{equation}
 \int^\infty_y {\rm d} y' \,\frac{1}{1+ {y'}^{6}} = \frac{1}{2 \sqrt{3}}
  \left[ \tan^{-1} (\sqrt{3} - 2 y) +\tan^{-1} (\sqrt{3} + 2 y)  \right] + \frac{1}{12} \log \left(\frac{1- y^2 + y^4}{1+2 y^2+y^4} \right),
\end{equation}
and
\begin{equation}
\begin{aligned}
 \int^\infty_y  {\rm d} y' \, & \frac{{y'}^2}{1+ {y'}^{10}}  = \frac{1}{10} \left[2 \tan^{-1}y -(1-\sqrt{5}) \pi \right] \\
&+ \frac{1}{20}
\left\{ (1-\sqrt{5}) \left[\tan^{-1}\left(\frac{\sqrt{10-2 \sqrt{5}}+4 y}{1+\sqrt{5}}\right)-  \tan^{-1} \left(\frac{\sqrt{10-2 \sqrt{5}}-4 y}{1+\sqrt{5}}\right) \right] \right. \\
& \hspace{15mm}+ \left. (1+\sqrt{5}) \left[\tan^{-1}\left(\frac{\sqrt{10-2 \sqrt{5}}+4 y}{1-\sqrt{5}}\right)-  \tan^{-1} \left(\frac{\sqrt{10-2 \sqrt{5}}-4 y}{1-\sqrt{5}}\right) \right] \right\} \\
&-\frac{1}{40}\left\{\sqrt{10+2\sqrt{5}} \log \left(\frac{2-y \sqrt{10-2 \sqrt{5}} +2 y^2}{2+y \sqrt{10-2 \sqrt{5}}+2 y^2}\right) \right.\\
&\hspace{15mm} \left. -\sqrt{10-2 \sqrt{5}} \log \left(\frac{2-y \sqrt{10+2\sqrt{5}} +2 y^2}{2+y \sqrt{10+2\sqrt{5}}+2 y^2} \right) \right\},
\end{aligned}
\end{equation}
respectively.  We use these analytical expressions in our implementation in \texttt{CosmoTherm} to compute the heating rates, and solve numerically only the $k$-independent viscosity component of the $k_D^{-2}$ differential equation~(\ref{eq:interpolatediffusion}).

\bibliography{ref}

\bibliographystyle{utcaps}

\end{document}